\documentclass[final,1p,times,number]{elsarticle}

\usepackage{amsmath}
\usepackage{amssymb}
\usepackage{amsfonts}
\usepackage{graphicx}
\usepackage{epstopdf}
\usepackage{color}
\usepackage{bm}
\usepackage{multirow}
\usepackage{natbib}
\usepackage{hyperref}
\usepackage{cleveref}

\usepackage{color}
\usepackage{float}
\usepackage{caption}

\usepackage[caption=false]{subfig}

\usepackage{ifpdf}
\ifpdf%
\usepackage{pdflscape}
\else
\usepackage{lscape}
\fi

\hypersetup{
	colorlinks,
	linkcolor={red!50!black},
	citecolor={blue!50!black},
	urlcolor={blue!80!black}
}

\def\e{\mbox{\boldmath $e$}}
\def\f{\mbox{\boldmath $f$}}

\def\g{\mbox{\boldmath $g$}}
\def\h{\mbox{\boldmath $h$}}
\def\m{\mbox{\boldmath $m$}}

\def\0{\mbox{\boldmath $0$}}

\biboptions{sort&compress}

\begin{document}

\begin{frontmatter}

\title{A Comparative Study of Projected and Unprojected Schemes for Micromagnetic Simulations}

\author[XJTLU]{Changjian Xie\corref{cor}}
\cortext[cor]{Corresponding author.} 
\ead{Changjian.Xie@xjtlu.edu.cn}



\address[XJTLU]{School of Mathematics and Physics, Xi'an-Jiaotong-Liverpool University, Re'ai Rd. 111, Suzhou, 215123, Jiangsu, China.}

\begin{abstract}

In micromagnetic simulations, the constant magnitude of the magnetization can be derived from the continuity equation. Since the time evolution of the magnetization in the continuity equation is perpendicular to the plane determined by the magnetization and the effective field, taking the inner product of both sides of the model with the magnetization shows that the evolution rate of the magnitude of the magnetization is zero, thus keeping the magnitude constant. From this perspective, the equation itself can maintain the constraint of constant magnetization magnitude. We discretized the continuity equation and compared two first-order semi-implicit strategies in time: one is the implicit Gauss-Seidel method, and the other is the semi-implicit Backward Differentiation Formula (BDF) method. We considered the comparison between these two schemes with and without the projection step. The results of micromagnetic simulations show that when the dissipation coefficient is large, the implicit Gauss-Seidel method without the projection step has significant differences from the method with the projection step in both the achieved steady state and domain wall motion. When an appropriate dissipation coefficient is selected, the difference between the two narrows, and both the steady state and domain wall motion can be simulated. For the other method, BDF1, whether the dissipation coefficient is large or small, the results with and without the projection step are quite consistent, and it can effectively simulate the domain wall motion.

\end{abstract}


\end{frontmatter}

\section{Introduction}

The temporal evolution of magnetization in ferromagnetic materials is fundamentally governed by the Landau-Lifshitz-Gilbert (LLG) equation \cite{Landau1935On,Gilbert:1955}. 
Such an equation forms a vector-valued nonlinear system subject to a pointwise constant-magnitude constraint on the magnetization vector. Extensive research on preserving the constraint has been devoted to developing efficient and numerically stable algorithms for micromagnetic simulations, see references \cite{kruzik2006recent,cimrak2007survey} and therein. Although numerous studies have investigated constraint-preserving algorithms for the LLG equation, see \cite{wang2001gauss}, \cite{qing_2023}, \cite{xiaoli_2024}, \cite{JEONG2010613}, and \cite{An2016OptimalEE}, such methods are not directly derived from the governing equation itself. Instead, starting from the requirement of norm preservation, they typically introduce nonlinear projection steps, adopt Lagrange multipliers, or embed the constraint indirectly into the discrete formulation.

Starting from the equation, in fact, we should not impose any constraints. We compared two typical methods: one is the Gauss-Seidel projection method, and the other is the semi-implicit projection method based on BDF1. Both types of methods apply a nonlinear projection step at the end, which brings additional difficulties to the mathematical analysis. From the perspective of engineering applications, the extent of the difference between the projection and non-projection methods in numerical simulations has not yet been studied. People usually consider designing numerical methods that satisfy the normalization condition, and it is also particularly important in applications to simply discretize the original continuity equation. In this paper, we present various results of micromagnetic simulations and apply them to domain wall motion. Starting from energy stability, we attempt to prove the energy stability of the non-projection method and simulate the energy curves under various dissipation conditions.

The rest of this paper is structured as follows. \Cref{sec: model} introduce the Landau-Lifshitz-Gilbert micromagnetic equation, then \Cref{sec: numerical scheme} provides two detailed formulation of the popular numerical schemes, which contains the first order Gauss-Seidel projection (no projection) method and the semi-implicit BDF1 projection (no projection) method. \Cref{sec:experiments} presents the simulation of micromagnetic statics and dynamics to study the domain wall motion.
Finally, concluding remarks and perspectives for future work are given in \Cref{sec:conclusions}.

\section{The LLG equation}
\label{sec: model}


The micromagnetic equation integrates gyromagnetic precession and dissipative relaxation \cite{Landau1935On,Brown1963micromagnetics}. Its nondimensionalized model is given by
\begin{align}\label{c1-large}
{\m}_t =-{\m}\times{\bm h}_{\text{eff}}-\alpha{\m}\times({\m}\times{\bm h}_{\text{eff}}),
\end{align}
where $\partial_t \m$ is the time derivative for the magnetization and ${\bm h}_{\text{eff}}$ is the term from the energy variation, say
\(\bm h_{\text{eff}} = -\delta E[\m]/\delta \m\). In this paper, we take the free energy below,
\begin{align*}
    E[\m]=\epsilon \int_{\Omega} |\nabla \m|^2\;dx+Q\int_{\Omega} (m_2^2+m_3^2)\;dx- \int_{\Omega} \h_s\cdot \m\;dx-\int_{\Omega}\h_e\cdot \m\;dx,
\end{align*}
(here $Q$ and $\epsilon$ are nondimensionalized parameters) and get the effective field below,
\begin{align*}
    {\bm h}_{\text{eff}}=\epsilon \Delta \m-Q(m_2\e_2+m_3\e_3)+\h_s+\h_e,
\end{align*}
where $\epsilon \Delta \m$ is the exchange field, $-Q(m_2\e_2+m_3\e_3)$ is anisotropy field, $\h_s$ is the stray field and $\h_e$ is the applied external field.

This equation is subject to the homogeneous Neumann boundary condition
\begin{equation}\label{boundary-large}
\frac{\partial{\m}}{\partial {\bm \nu}}\Big|_{\partial \Omega}=0,
\end{equation}
where \(\Omega \subset \mathbb{R}^d\) (\(d=1,2,3\)) denotes the bounded spatial domain occupied by the ferromagnetic material, and \(\bm \nu\) represents the unit outward normal vector on the domain boundary \(\partial \Omega\). It is noteworthy that this boundary condition is physically consistent for isolated ferromagnetic systems, as it inherently ensures the absence of magnetic surface charge—an essential prerequisite for accurately modeling unperturbed magnetic dynamics.

To achieve a comprehensive understanding of the LLG equation, it is essential to elucidate the physical essence of its key components. The magnetization field \(\m: \Omega \to \mathbb{R}^3\) is a three-dimensional vector field subject to the pointwise constraint \(|\m|=1\)—a fundamental characteristic derived from the quantum mechanical alignment of electron spins in ferromagnetic materials. With respect to the right-hand side of \cref{c1-large}, the first term describes the gyromagnetic precession effect, whereby magnetic moments undergo precessional motion around the effective magnetic field \(\bm h_{\text{eff}}\). The second term represents dissipative relaxation, where the parameter \(\alpha > 0\) stands for the dimensionless Gilbert damping coefficient, which quantifies the rate of energy transfer from the magnetic subsystem to the lattice structure. Consequently, the model \cref{c1-large} inherently satisfies the norm-preserving constraint.

\section{Typical methods with projection}\label{sec: numerical scheme}

\subsection{Gauss-Seidel projection method}
The Gauss-Seidel projection method (GSPM) is given as the following three steps:

\paragraph{Step 1. Implicit Gauss-Seidel:}
\begin{align}
g_i^n &= (I - \epsilon \Delta t \Delta_h)^{-1} (m_i^n + \Delta t f_i^n), \\
g_i^* &= (I - \epsilon \Delta t \Delta_h)^{-1} (m_i^* + \Delta t f_i^n), \quad i=1,2,3
\end{align}
\begin{equation}
\begin{pmatrix}
m_1^* \\
m_2^* \\
m_3^*
\end{pmatrix}
=
\begin{pmatrix}
m_1^n + (g_2^n m_3^n - g_3^n m_2^n) \\
m_2^n + (g_3^n m_1^* - g_1^* m_3^n) \\
m_3^n + (g_1^* m_2^* - g_2^* m_1^*)
\end{pmatrix}.
\label{eq:step1_update}
\end{equation}

\paragraph{Step 2. Heat flow without constraints:}
\begin{equation}
\boldsymbol{f}^* = -Q(m_2^* \boldsymbol{e}_2 + m_3^* \boldsymbol{e}_3) + \boldsymbol{h}_s^n + \boldsymbol{h}_e,
\label{eq:heat_flow_f}
\end{equation}
\begin{equation}
\begin{pmatrix}
m_1^{**} \\
m_2^{**} \\
m_3^{**}
\end{pmatrix}
=
\begin{pmatrix}
m_1^* + \alpha \Delta t (\epsilon \Delta_h m_1^{**} + f_1^*) \\
m_2^* + \alpha \Delta t (\epsilon \Delta_h m_2^{**} + f_2^*) \\
m_3^* + \alpha \Delta t (\epsilon \Delta_h m_3^{**} + f_3^*)
\end{pmatrix}.
\label{eq:heat_flow_update}
\end{equation}

\paragraph{Step 3. Projection onto $S^2$:}
\begin{equation}
\begin{pmatrix}
m_1^{n+1} \\
m_2^{n+1} \\
m_3^{n+1}
\end{pmatrix}
=
\frac{1}{|m^{**}|}
\begin{pmatrix}
m_1^{**} \\
m_2^{**} \\
m_3^{**}
\end{pmatrix}.
\label{eq:projection}
\end{equation}

where $\boldsymbol{m}^*$ denotes the intermediate values of $\boldsymbol{m}$. The stray field $\boldsymbol{h}_s$ is computed using the intermediate values $\boldsymbol{m}^*$ in \eqref{eq:step1_update} and \eqref{eq:heat_flow_f}.

\subsection{Semi-implicit projection methods}
The BDF1 projection method is given below,
\begin{align*}
    \left\{
\begin{aligned}
\frac{\tilde{\boldsymbol{m}}_h^{n+1} - \boldsymbol{m}_h^n}{k}
&= -\boldsymbol{m}_h^n \times \left( \epsilon \Delta_h \tilde{\boldsymbol{m}}_h^{n+1} + \boldsymbol{f}_h^n \right) \\
&\quad - \alpha \boldsymbol{m}_h^n \times \left( \boldsymbol{m}_h^n \times \left( \epsilon \Delta_h \tilde{\boldsymbol{m}}_h^{n+1} + \boldsymbol{f}_h^n \right) \right), \\
\boldsymbol{m}_h^{n+1} &= \frac{\tilde{\boldsymbol{m}}_h^{n+1}}{\left| \tilde{\boldsymbol{m}}_h^{n+1} \right|},
\end{aligned}
\right.
\end{align*}

\section{Typical methods without projection}

\subsection{Gauss-Seidel without projection method}
The Gauss-Seidel without projection method (GSnoPM) is given as the following three steps:

\paragraph{Step 1. Implicit Gauss-Seidel:}
\begin{align}
g_i^n &= (I - \epsilon \Delta t \Delta_h)^{-1} (m_i^n + \Delta t f_i^n), \\
g_i^* &= (I - \epsilon \Delta t \Delta_h)^{-1} (m_i^* + \Delta t f_i^n), \quad i=1,2,3
\end{align}
\begin{equation}
\begin{pmatrix}
m_1^* \\
m_2^* \\
m_3^*
\end{pmatrix}
=
\begin{pmatrix}
m_1^n + (g_2^n m_3^n - g_3^n m_2^n) \\
m_2^n + (g_3^n m_1^* - g_1^* m_3^n) \\
m_3^n + (g_1^* m_2^* - g_2^* m_1^*)
\end{pmatrix}.
\label{eq:step1_update}
\end{equation}

\paragraph{Step 2. Heat flow without constraints:}
\begin{equation}
\boldsymbol{f}^* = -Q(m_2^* \boldsymbol{e}_2 + m_3^* \boldsymbol{e}_3) + \boldsymbol{h}_s^n + \boldsymbol{h}_e,
\label{eq:heat_flow_f}
\end{equation}
\begin{equation}
\begin{pmatrix}
m_1^{n+1} \\
m_2^{n+1} \\
m_3^{n+1}
\end{pmatrix}
=
\begin{pmatrix}
m_1^* + \alpha \Delta t (\epsilon \Delta_h m_1^{n+1} + f_1^*) \\
m_2^* + \alpha \Delta t (\epsilon \Delta_h m_2^{n+1} + f_2^*) \\
m_3^* + \alpha \Delta t (\epsilon \Delta_h m_3^{n+1} + f_3^*)
\end{pmatrix}.
\label{eq:heat_flow_update}
\end{equation}


where $\boldsymbol{m}^*$ denotes the intermediate values of $\boldsymbol{m}$. The stray field $\boldsymbol{h}_s$ is computed using the intermediate values $\boldsymbol{m}^*$ in \eqref{eq:step1_update} and \eqref{eq:heat_flow_f}.

For brevity, we take $\epsilon=1$ and $\f_h^n=0$ and for the step 1, we note that
\begin{align}\label{eq-s}
    \frac{\m_h^{*}-\m_h^n}{k}=-\m_h^s\times \Delta_h \g_h^s,
\end{align}
where $\g_h^s=(I-k\Delta_h)^{-1}\m_h^{s}$, $s=n,*$. The linear stability analysis can not work. However, we notice that for Gauss-Seidel iteration method,
\begin{align}\label{gspm-m}
    \begin{pmatrix}
        1&&\\
        -g_3^n&1&\\
        g_2^*&-g_1^*&1
    \end{pmatrix}\begin{pmatrix}
        m_1^*\\
        m_2^*\\
        m_3^*
    \end{pmatrix}=\begin{pmatrix}
        1&-g_3^n&g_2^n\\
        &1&-g_1^*\\
        &&1
    \end{pmatrix}\begin{pmatrix}
        m_1^n\\
        m_2^n\\
        m_3^n
    \end{pmatrix}
\end{align}
If $\g$ is a constant vector, we can do the spectral analysis for matrix $A$ of $\m_h^{n+1}=A \m_h^n$. However, here, the matrix depends on the previous $\m_h^{*}$. The exact stability property of the scheme \cref{gspm-m} is difficult to analyze. Numerical experimentation suggests that the scheme is unconditionally stable.


Noting that 
\begin{align*}
    \frac{g_3^n-m_3^n}{k}=\Delta_h g_3^n,\quad \frac{g_2^n-m_2^n}{k}=\Delta_h g_2^n,
\end{align*}
We can derive that $\|\nabla_h g_3^n\|_2^2\le \|\nabla_h m_3^n\|_2^2$ and $\|\nabla_h g_2^n\|_2^2\le \|\nabla_h m_2^n\|_2^2$. Similarly, we have
\begin{align*}
    \|\nabla_h g_1^*\|_2^2\le \|\nabla_h m_1^*\|_2^2,\quad  \|\nabla_h g_2^*\|_2^2\le \|\nabla_h m_2^*\|_2^2.
\end{align*}
say, we have $\|\nabla_h \g_h^s\|_2^2\le \|\nabla_h m_h^s\|_2^2$.
We take inner product with $\Delta_h \g_h^s$, we get 
\begin{align*}
    RHS=(-\m_h^s\times \Delta_h \g_h^s,\Delta_h \g_h^s)=0,
\end{align*}
and 
\begin{align*}
    LHS=\frac{1}{k} (\m_h^{*}-\m_h^n,\Delta_h \g_h^s)=-\frac{1}{k}(\nabla_h\m_h^{*}-\nabla_h\m_h^n,\nabla_h \g_h^s),
\end{align*}
Thus,
we can prove that 
\begin{align*}
    \|\nabla_h \m_h^{*}\|_2^2 \le  \|\nabla_h \m_h^{n}\|_2^2.
\end{align*}

For step 2, we have
\begin{align*}
    \frac{\m_h^{n+1}-\m_h^{*}}{k}=\alpha \Delta_h \m_h^{n+1}.
\end{align*}
We take the inner product with $-\Delta_h \m_h^{n+1}$, and have
\begin{align*}
    LHS&=\frac{1}{k}(\m_h^{n+1}-\m_h^{*},-\Delta_h \m_h^{n+1})\\
    &=\frac{1}{2k}\left(\|\nabla_h \m_h^{n+1}\|_2^2+\|\nabla_h \m_h^{n+1}-\nabla_h \m_h^{*}\|_2^2-\|\nabla_h \m_h^{*}\|_2^2 \right),
\end{align*}
and 
\begin{align*}
    RHS=-\alpha(\Delta_h \m_h^{n+1},\Delta_h \m_h^{n+1})=-\alpha\|\Delta_h \m_h^{n+1}\|_2^2 \le 0.
\end{align*}
We thus have
\begin{align*}
    \|\nabla_h \m_h^{n+1}\|_2^2 \le  \|\nabla_h \m_h^{*}\|_2^2.
\end{align*}

\subsection{Semi-implicit without projection method}
The BDF1 without projection method is given below,
\begin{align}\label{eq-BDF1}
\frac{{\boldsymbol{m}}_h^{n+1} - \boldsymbol{m}_h^n}{k}
&= -\boldsymbol{m}_h^n \times \left( \epsilon \Delta_h {\boldsymbol{m}}_h^{n+1} + \boldsymbol{f}_h^n \right) - \alpha \boldsymbol{m}_h^n \times \left( \boldsymbol{m}_h^n \times \left( \epsilon \Delta_h {\boldsymbol{m}}_h^{n+1} + \boldsymbol{f}_h^n \right) \right). 
\end{align}

For such a BDF1 method, we can give a stability analysis below, for brevity, we take $\epsilon=1$ and $\f_h^n=0$ and take inner product with $-\Delta_h \m_h^{n+1}$, and obtain
\begin{align*}
    LHS&=\frac{1}{k}(\m_h^{n+1}-\m_h^n,-\Delta_h \m_h^{n+1})\\
    &=\frac{1}{2k}\left(\|\nabla_h \m_h^{n+1}\|_2^2+\|\nabla_h \m_h^{n+1}-\nabla_h \m_h^n\|_2^2-\|\nabla_h \m_h^n\|_2^2 \right),
\end{align*}
and
\begin{align*}
    RHS&=\alpha \left(\Delta_h \m_h^{n+1},\m_h^n\times (\m_h^n\times \Delta_h \m_h^{n+1}) \right)\\
    &=\alpha \left(\m_h^n\times \Delta_h \m_h^{n+1},\Delta_h \m_h^{n+1}\times \m_h^n \right)=-\alpha \|\m_h^n\times \Delta_h \m_h^{n+1}\|_2^2\le 0.
\end{align*}
We then have
\begin{align*}
    \|\nabla_h \m_h^{n+1}\|_2^2 \le  \|\nabla_h \m_h^{n}\|_2^2.
\end{align*}

If $\f^n_h=-Q(m_2^n\e_2+m_3^n \e_3)+\h_s(\m_h^n)+\h_e$, we take the inner product for \cref{eq-BDF1} with $-\epsilon \Delta_h \m_h^{n+1}+\f_h^n$, and we obtain that
\begin{align*}
    LHS&=\frac{1}{k}(\m_h^{n+1}-\m_h^n,-\epsilon \Delta_h \m_h^{n+1}+\f_h^n)\\
    &=\frac{1}{2k}\left(\epsilon\|\nabla_h \m_h^{n+1}\|_2^2+\epsilon\|\nabla_h \m_h^{n+1}-\nabla_h \m_h^n\|_2^2-\epsilon\|\nabla_h \m_h^n\|_2^2 +2(\m_h^{n+1},\f_h^n)-2(\m_h^n,\f_h^n)\right),
\end{align*}
and 
\begin{align*}
    RHS&=\alpha \left(\epsilon\Delta_h \m_h^{n+1}+\f_h^n,\m_h^n\times (\m_h^n\times (\epsilon\Delta_h \m_h^{n+1}+\f_h^n)) \right)\\
    &=\alpha \left(\m_h^n\times (\epsilon\Delta_h \m_h^{n+1}+\f_h^n)),(\epsilon\Delta_h \m_h^{n+1}+\f_h^n))\times \m_h^n \right)=-\alpha \|\m_h^n\times (\epsilon\Delta_h \m_h^{n+1}+\f_h^n))\|_2^2\le 0.
\end{align*}
Thus, we have
\begin{align*}
    \|\nabla_h \m_h^{n+1}\|_2^2 +2(\f_h^n,\m_h^{n+1})\le  \|\nabla_h \m_h^{n}\|_2^2+2(\f_h^n,\m_h^{n}).
\end{align*}
The BDF1 method without projection is energy stable in the discrete level.

\section{Micromagnetics simulations}\label{sec:experiments}

\begin{figure}[htbp]
    \centering
    \subfloat[Initial S state arrow]{\includegraphics[width=0.3\linewidth]{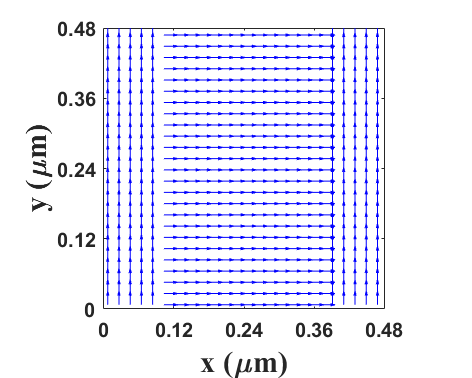}}
     \subfloat[GSPM projection arrow]{\includegraphics[width=0.3\linewidth]{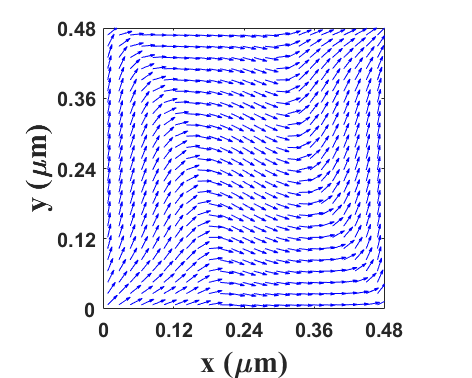}}
     \subfloat[GSPM no projection arrow]{\includegraphics[width=0.3\linewidth]{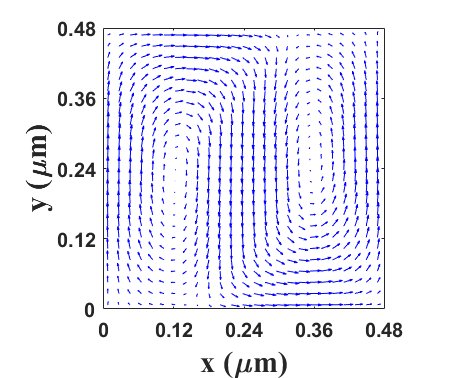}}
    \hspace{0.1in}
     \subfloat[Initial S state angle]{\includegraphics[width=0.3\linewidth]{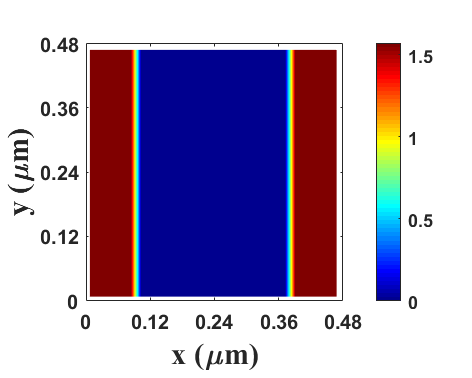}}
      \subfloat[GSPM projection angle]{\includegraphics[width=0.3\linewidth]{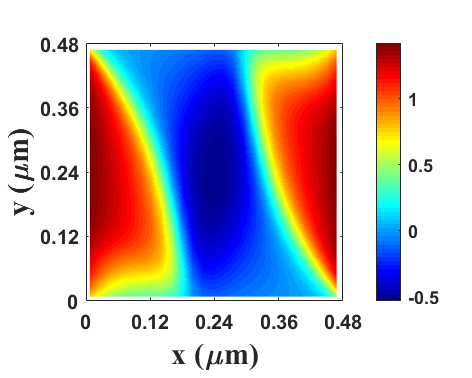}}
      \subfloat[GSPM noprojection angle]{\includegraphics[width=0.3\linewidth]{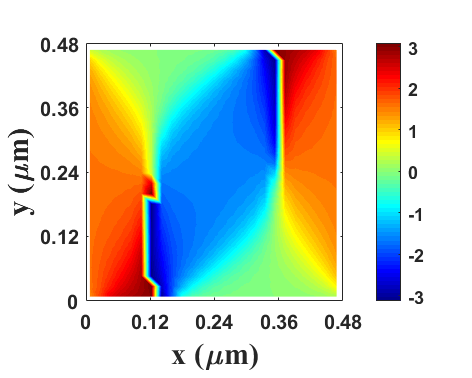}}
     \hspace{0.1in}
      \subfloat[Initial S state arrow]{\includegraphics[width=0.3\linewidth]{gspm_noprojection_arrow_S_initial}}
     \subfloat[GSPM projection arrow]{\includegraphics[width=0.3\linewidth]{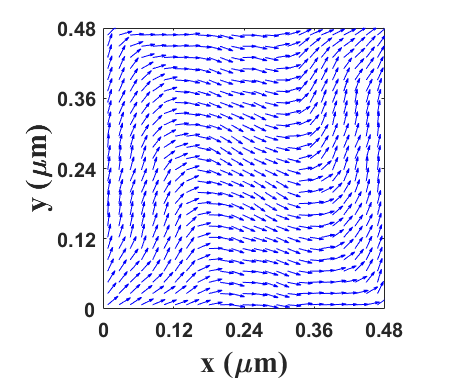}}
     \subfloat[GSPM no projection arrow]{\includegraphics[width=0.3\linewidth]{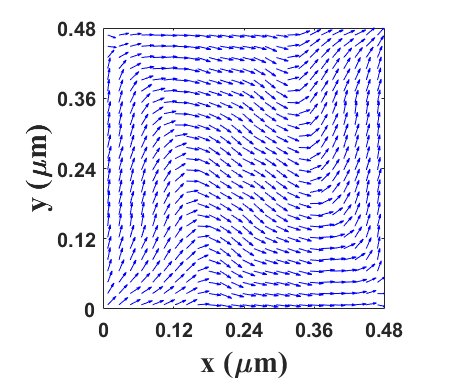}}
     \hspace{0.1in}
      \subfloat[Initial S state angle]{\includegraphics[width=0.3\linewidth]{gspm_noprojection_color_S_initial}}
      \subfloat[GSPM projection angle]{\includegraphics[width=0.3\linewidth]{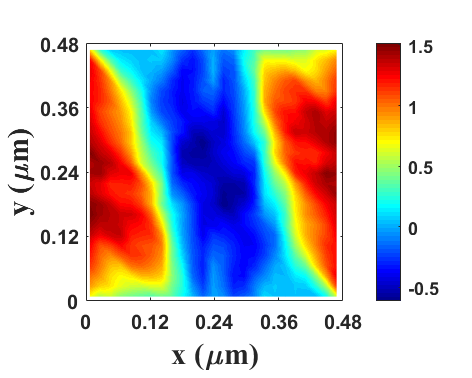}}
      \subfloat[GSPM noprojection angle]{\includegraphics[width=0.3\linewidth]{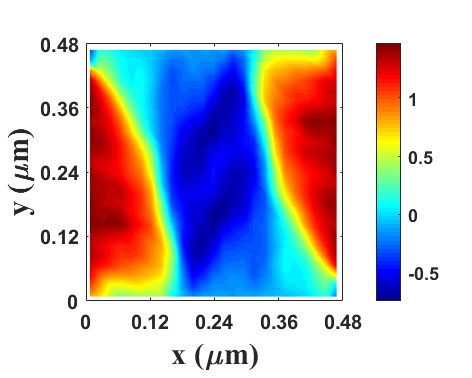}}
    \caption{The setup $\alpha=0.1$ (top two rows), $\alpha=0.01$ (bottom two rows) and the final time $1\;ns$. Initial state is given by left panel. The results by GSPM projection in middle panel. The results by GSPM no projection in middle panel.}
    \label{fig:1}
\end{figure}

\begin{figure}[htbp]
    \centering
    \subfloat[Initial S state arrow]{\includegraphics[width=0.3\linewidth]{gspm_noprojection_arrow_S_initial}}
     \subfloat[BDF1 projection arrow]{\includegraphics[width=0.3\linewidth]{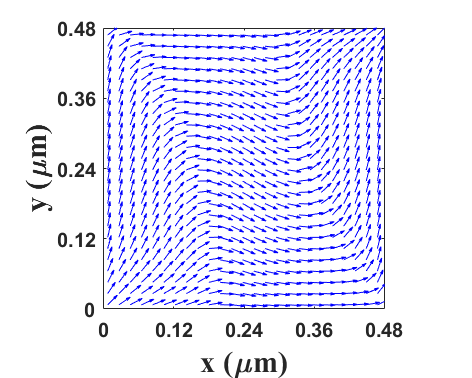}}
     \subfloat[BDF1 no projection arrow]{\includegraphics[width=0.3\linewidth]{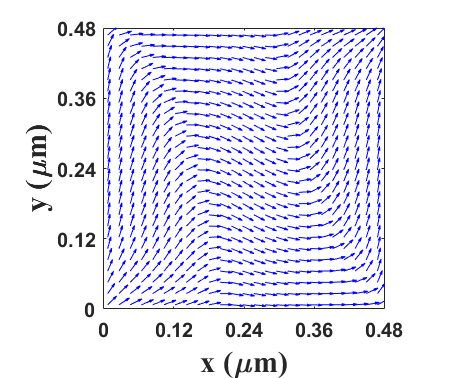}}
    \hspace{0.1in}
     \subfloat[Initial S state angle]{\includegraphics[width=0.3\linewidth]{gspm_noprojection_color_S_initial}}
      \subfloat[BDF1 projection angle]{\includegraphics[width=0.3\linewidth]{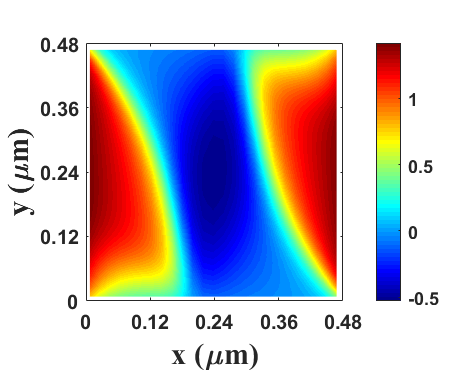}}
      \subfloat[BDF1 noprojection angle]{\includegraphics[width=0.3\linewidth]{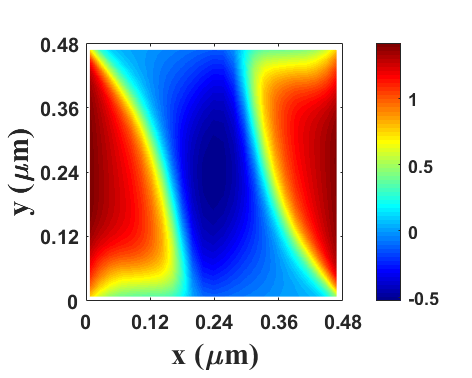}}
       \hspace{0.1in}
       \subfloat[Initial S state arrow]{\includegraphics[width=0.3\linewidth]{gspm_noprojection_arrow_S_initial}}
     \subfloat[BDF1 projection arrow]{\includegraphics[width=0.3\linewidth]{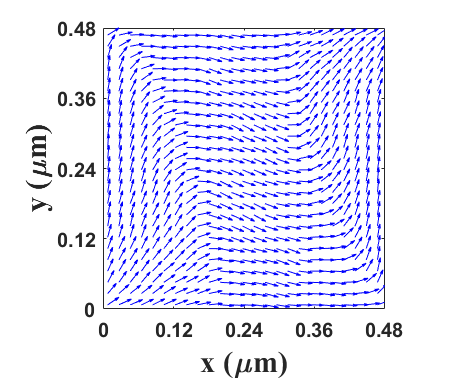}}
     \subfloat[BDF1 no projection arrow]{\includegraphics[width=0.3\linewidth]{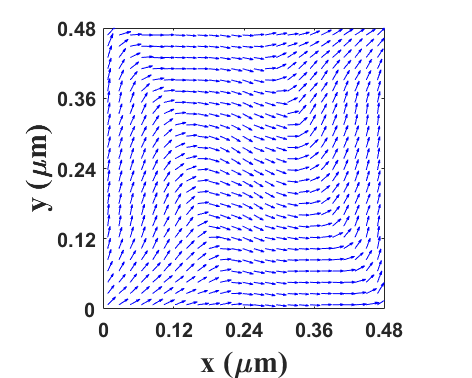}}
    \hspace{0.1in}
     \subfloat[Initial S state angle]{\includegraphics[width=0.3\linewidth]{gspm_noprojection_color_S_initial}}
      \subfloat[BDF1 projection angle]{\includegraphics[width=0.3\linewidth]{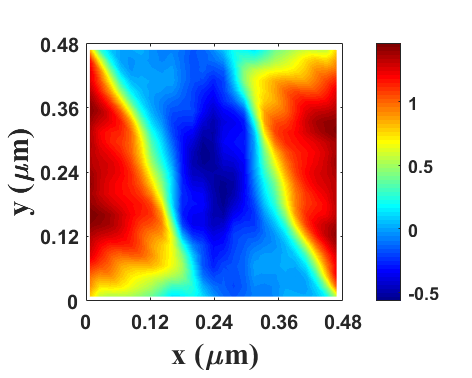}}
      \subfloat[BDF1 noprojection angle]{\includegraphics[width=0.3\linewidth]{BDF1_E_noprojection_color_S_alpha_0dot01_1ns}}
    \caption{The setup $\alpha=0.1$ (top two rows), $\alpha=0.01$ (bottom two rows) and the final time $1\;ns$. Initial state is given by left panel. The results by BDF1 projection in middle panel. The results by BDF1 no projection in middle panel.}
    \label{fig:BDF1-1}
\end{figure}

In this section, we first simulate the statics with different initial conditions by GSPM (with or without projection) method and BDF1 (with or without projection) method. The simulation model uses a ferromagnetic thin film with dimensions $480\times 480\times 20\;nm^3$ discretized on a $100\times 100 \times 4$ grid. The temporal step size we take is $k=1\;ps$. To compare the performance of different methods, we first pick up the initial state of S profile, the result of GSPM method with and without projection for the equilibrium at final time $T=1\;ns$ is presented in \Cref{fig:1}. It turns out that the GSPM without projection get a different profile than that of projection method with slightly large $\alpha=0.1$, however, the situation is better when slightly small $\alpha=0.01$. To verify such an observation, we give different initial state setup in the Appendix A. We choose C state, Diamond state, Flower state, random state, Single Crosstie state, Double Crosstie state for the initialization with $\alpha=0.1$ and get a similar observation. However, the results for BDF1 with projection and without projection are presented in \Cref{fig:BDF1-1}. It indicates that the equilibrium is comparable using BDF1 projection and no projection with a great tolerance of $\alpha=0.1$ and $\alpha=0.01$. To verify such an observation, we give different initial state setup in the Appendix B. We choose C state, Diamond state, Flower state, random state, Single Crosstie state, Double Crosstie state for the initialization with $\alpha=0.1$ and get a similar observation. From the perspective of statics, BDF1 is much more stable scheme than GSPM. We can use BDF1 no projection method to do the simulation for different $\alpha$, however, we may use GSPM without projection method to replace GSPM projection results with a slightly small $\alpha$. In short, the scheme without projection can be applied to the simulation of micormagnetics.

\begin{figure}[htbp]
    \centering
    \subfloat[gspm projection]{\includegraphics[width=0.5\linewidth]{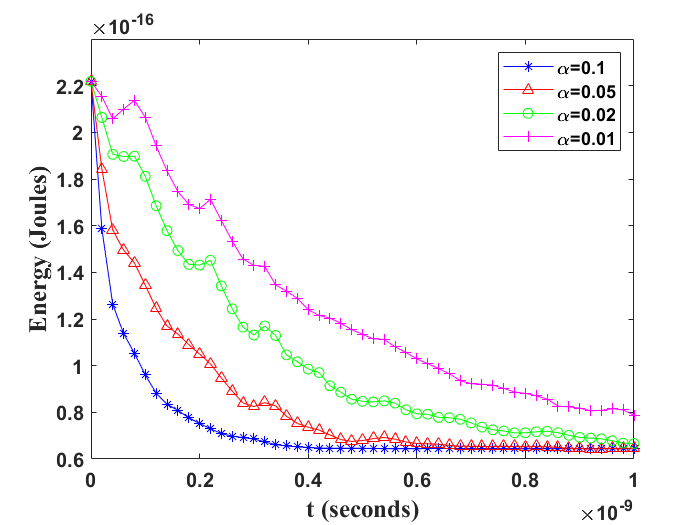}}
    \subfloat[gspm no projection]{\includegraphics[width=0.5\linewidth]{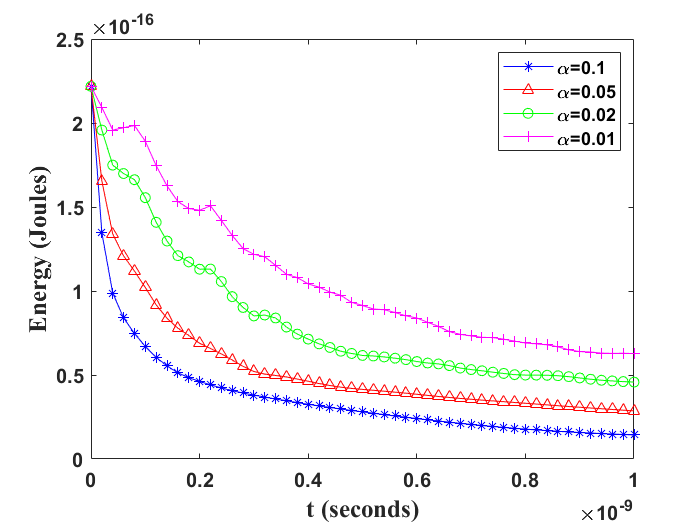}}
  \caption{The energy evolution over time for GSPM. The final time is $1\;ns$.}
    \label{fig:energy-1}
\end{figure}

\begin{figure}[htbp]
    \centering
    \subfloat[$\alpha=0.1$]{\includegraphics[width=0.5\linewidth]{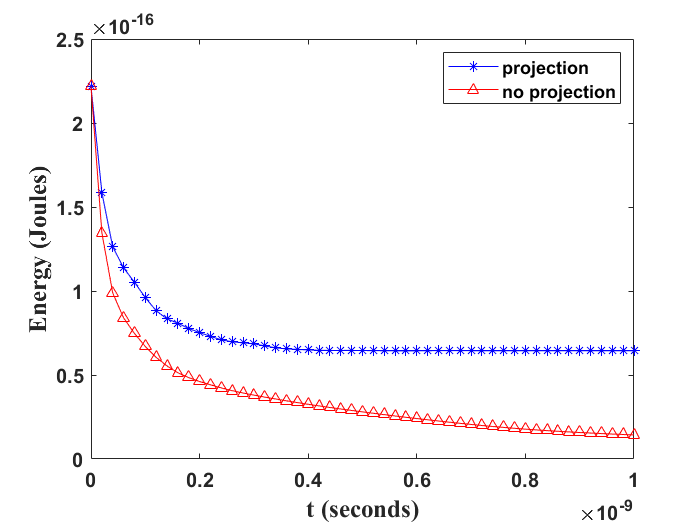}}
    \subfloat[$\alpha=0.05$]{\includegraphics[width=0.5\linewidth]{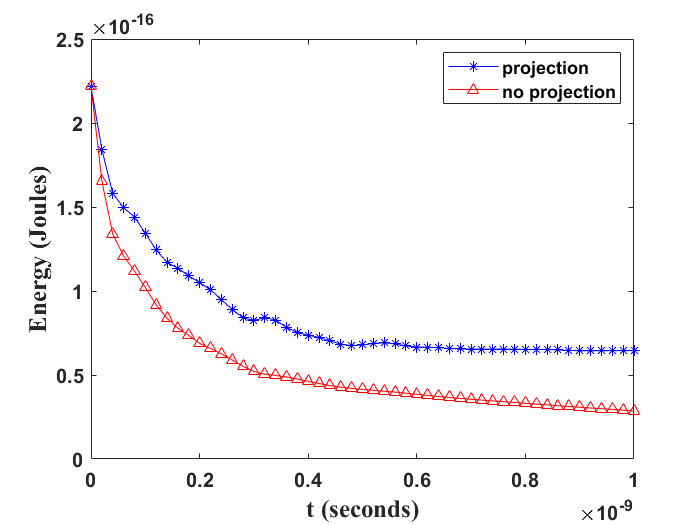}}
    \hspace{0.1in}
     \subfloat[$\alpha=0.02$]{\includegraphics[width=0.5\linewidth]{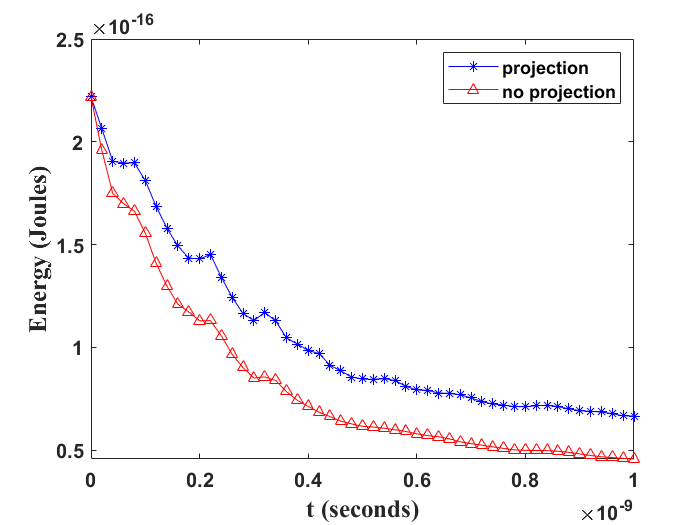}}
    \subfloat[$\alpha=0.01$]{\includegraphics[width=0.5\linewidth]{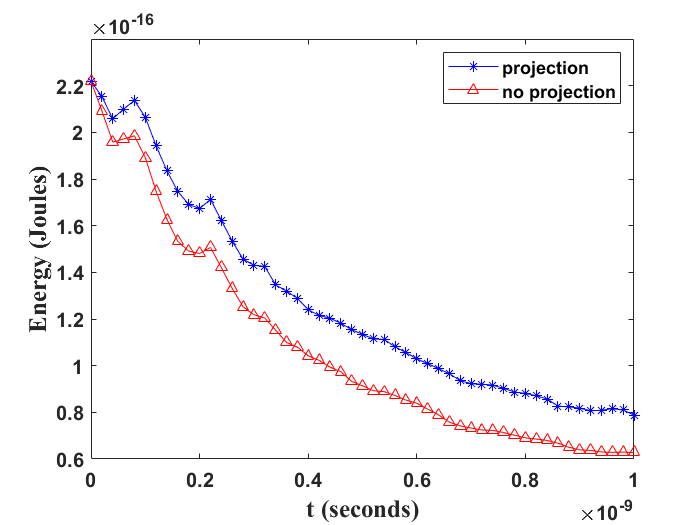}}
  \caption{The energy evolution over time for GSPM. The final time is $1\;ns$. For all $\alpha$, the energy level without projection is lower than that of projection method.}
    \label{fig:energy-2}
\end{figure}

\begin{figure}[htbp]
    \centering
    \subfloat[BDF1 projection]{\includegraphics[width=0.5\linewidth]{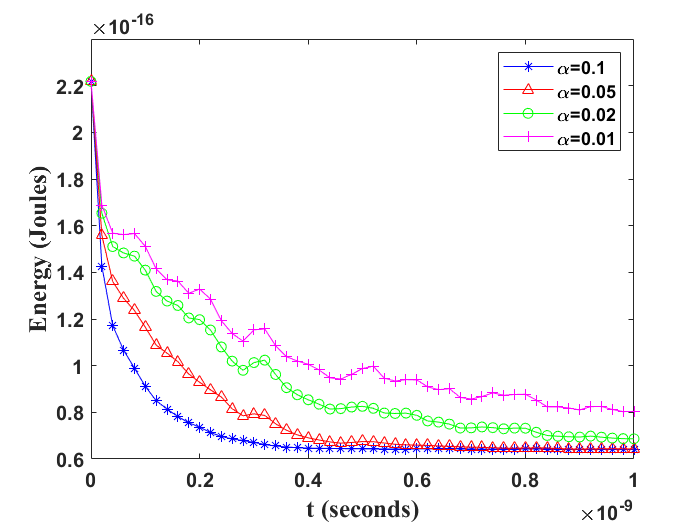}}
    \subfloat[BDF1 no projection]{\includegraphics[width=0.5\linewidth]{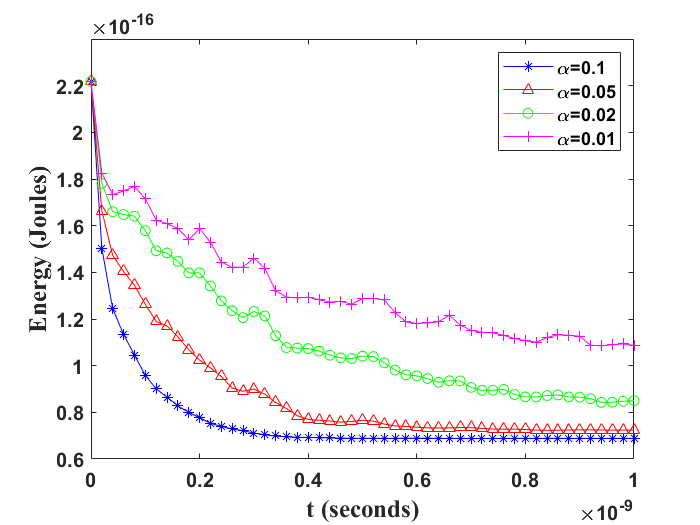}}
  \caption{The energy evolution over time for BDF1. The final time is $1\;ns$.}
    \label{fig:energy-BDF-1}
\end{figure}

\begin{figure}[htbp]
    \centering
    \subfloat[$\alpha=0.1$]{\includegraphics[width=0.5\linewidth]{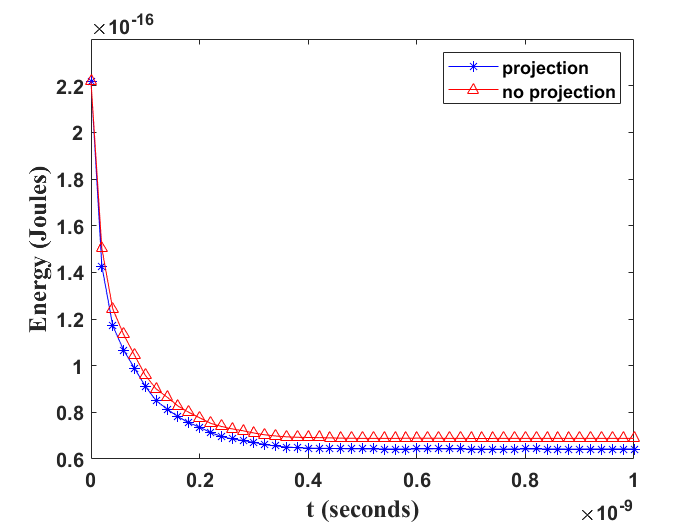}}
    \subfloat[$\alpha=0.05$]{\includegraphics[width=0.5\linewidth]{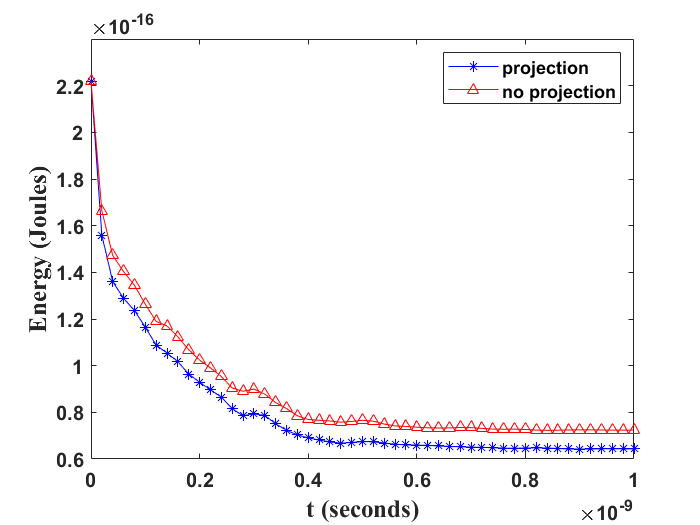}}
    \hspace{0.1in}
     \subfloat[$\alpha=0.02$]{\includegraphics[width=0.5\linewidth]{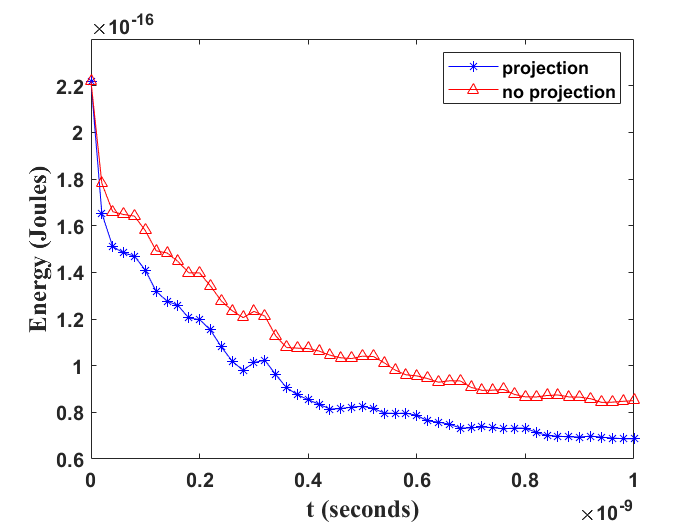}}
    \subfloat[$\alpha=0.01$]{\includegraphics[width=0.5\linewidth]{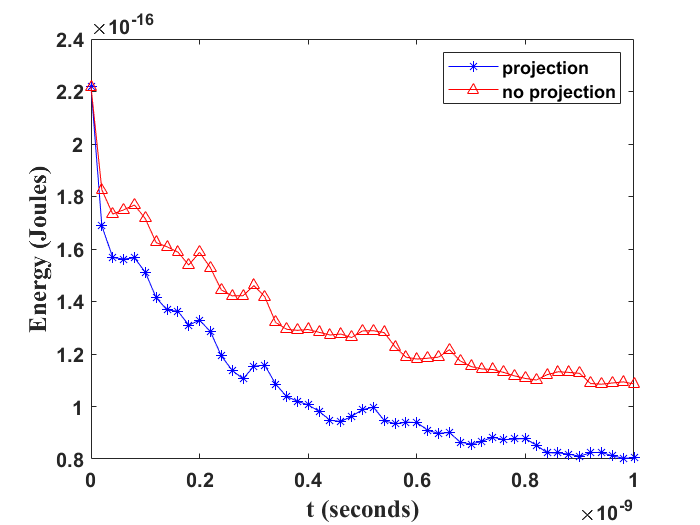}}
  \caption{The energy evolution over time for BDF1. The final time is $1\;ns$. For all $\alpha$, the energy level without projection is lower than that of projection method.}
    \label{fig:energy-BDF-2}
\end{figure}

To get the energy evolution over time, we set the initial condition as S state, and choose different damping parameters, say $\alpha=0.1, 0.05, 0.02, 0.01$. The result for GSPM with projection is presented in \Cref{fig:energy-1}(a). The result for GSPM without projection is presented in \Cref{fig:energy-1}(b). It suggests that the energy decreasing rate is faster as the damping parameter $\alpha$ larger for both situations. We take $\alpha=0.1, 0.05, 0.02, 0.01$ to compare the energy evolution using GSPM projection and no projection methods. \Cref{fig:energy-2} shows that the energy decreasing rate is faster using GSPM wihtout projection than that of GSPM projection method. The energy decreasing rate over time using BDF1 projection method and no projection method is presented in \Cref{fig:energy-BDF-1}(a) and \Cref{fig:energy-BDF-1}(b). It turns out the deacrasing rate is faster as the damping parameter $\alpha$ larger for both situations. The energy decreasing rate is faster using BDF1 projection method than that of BDF1 no projection method, which is shown in \Cref{fig:energy-BDF-2}. The results for BDF1 projection and no projection method in \Cref{fig:energy-BDF-2} is deferent with GSPM projection and no projection method, shown in \Cref{fig:energy-2}.

\begin{figure}[htbp]
    \centering
    \subfloat[$\alpha=0.1$]{\includegraphics[width=0.35\linewidth]{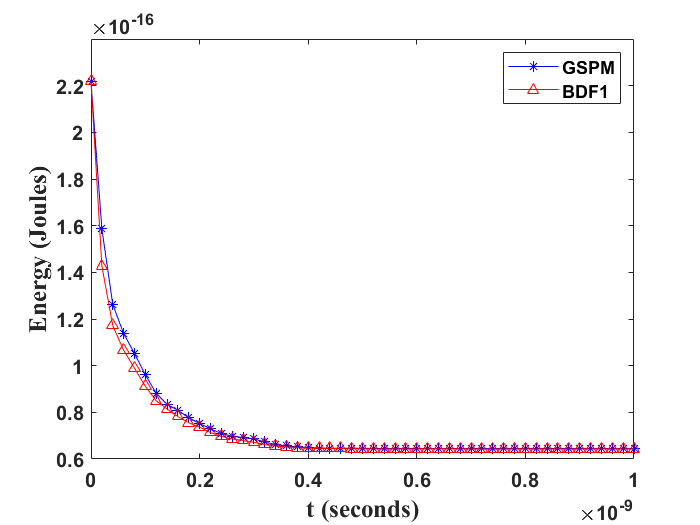}}
    \subfloat[$\alpha=0.05$]{\includegraphics[width=0.35\linewidth]{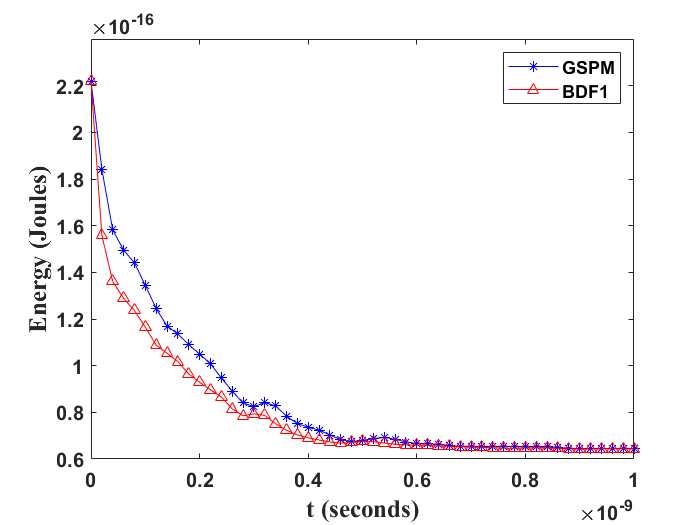}}
    \hspace{0.1in}
     \subfloat[$\alpha=0.02$]{\includegraphics[width=0.35\linewidth]{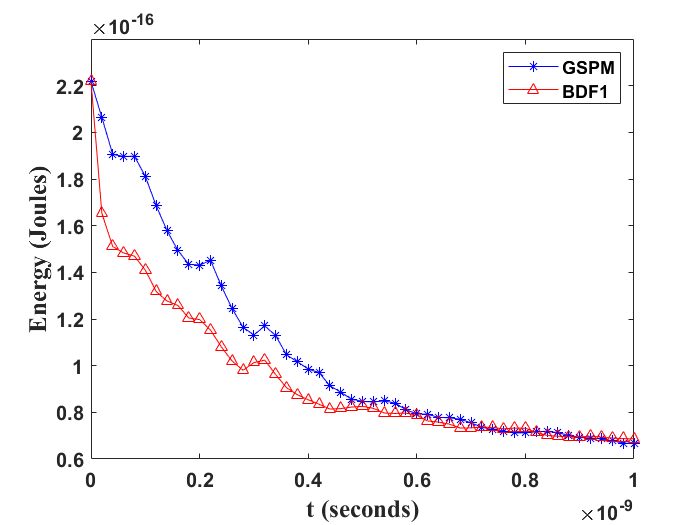}}
    \subfloat[$\alpha=0.01$]{\includegraphics[width=0.35\linewidth]{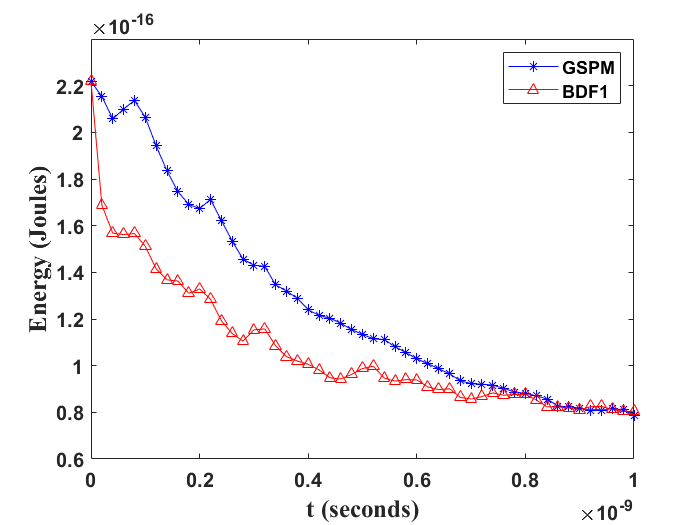}}
  \caption{The energy evolvement over time for BDF1 and GSPM with projection. The final time is $1\;ns$. }
    \label{fig:energy-BDF1-gspm-1}
\end{figure}

\begin{figure}[htbp]
    \centering
    \subfloat[$\alpha=0.1$]{\includegraphics[width=0.35\linewidth]{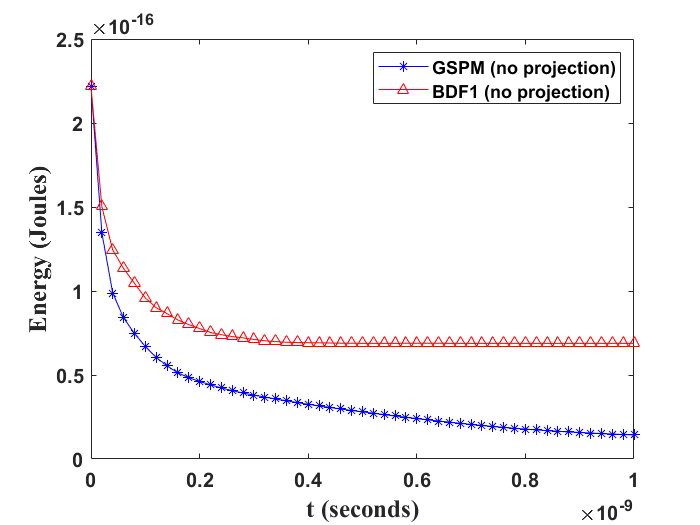}}
    \subfloat[$\alpha=0.05$]{\includegraphics[width=0.35\linewidth]{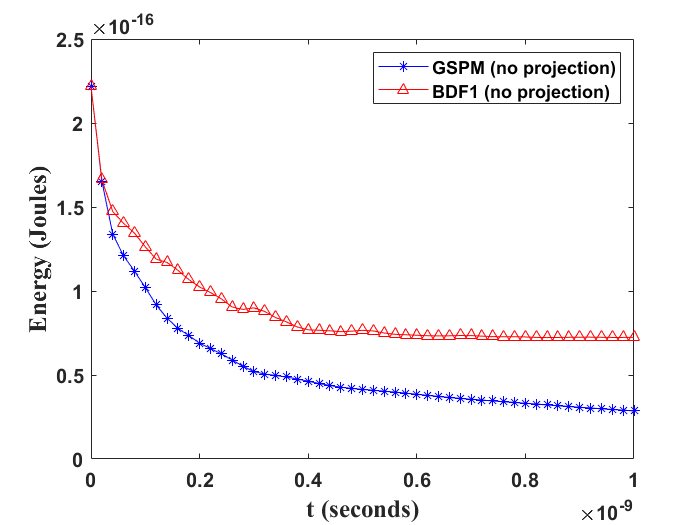}}
    \hspace{0.1in}
     \subfloat[$\alpha=0.02$]{\includegraphics[width=0.35\linewidth]{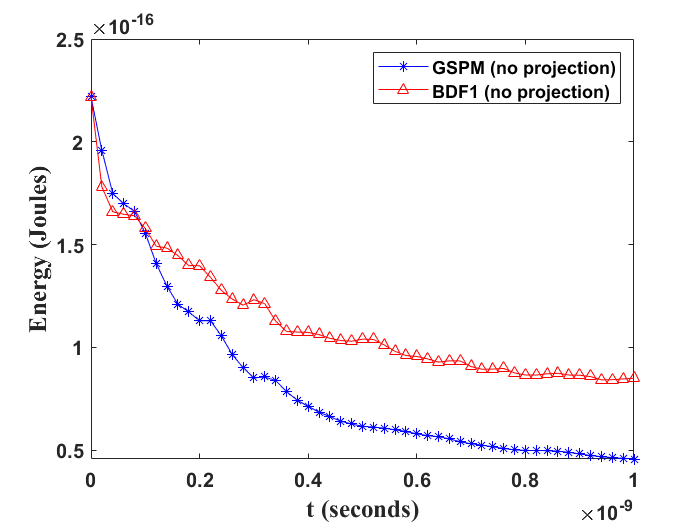}}
    \subfloat[$\alpha=0.01$]{\includegraphics[width=0.35\linewidth]{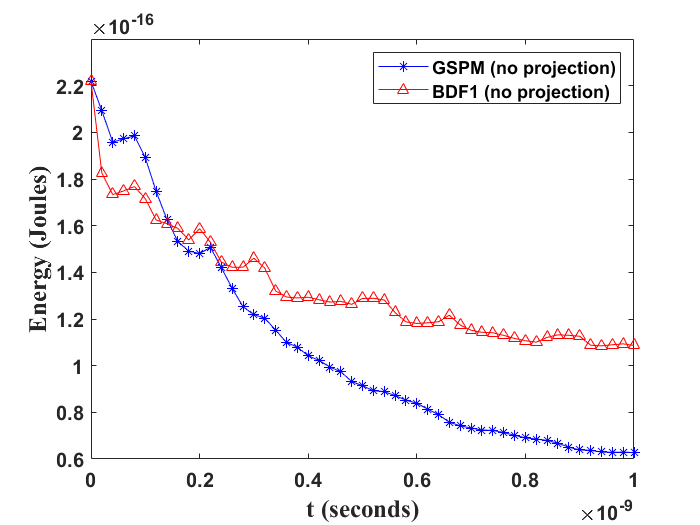}}
  \caption{The energy evolvement over time for BDF1 and GSPM without projection. The final time is $1\;ns$. }
    \label{fig:energy-BDF1-gspm-2}
\end{figure}

To better compare the energy difference between BDF1 and GSPM with or without projection, we obtain the results presented in \Cref{fig:energy-BDF1-gspm-1} and \Cref{fig:energy-BDF1-gspm-2}. As illustrated in the \Cref{fig:energy-BDF1-gspm-1}, we compare the temporal evolution of energy (up to 1 ns) between the GSPM and BDF1 methods with projection under four distinct $\alpha$ values (0.1, 0.05, 0.02, and 0.01). The results demonstrate that the energy profiles of the two methods are nearly identical and stable for relatively large $\alpha$ (0.1 and 0.05). As $\alpha$ decreases, the GSPM method shows increasingly pronounced energy oscillations, whereas the BDF1 method maintains a smooth and monotonic energy decay. Notably, for small $\alpha$ values (0.02 and particularly 0.01), GSPM suffers from severe energy fluctuations and departs from the stable energy decay trend, while BDF1 remains robust, indicating that BDF1 outperforms GSPM in maintaining reliable energy behavior and exhibits stronger numerical stability for small $\alpha$ values.

As depicted in the \Cref{fig:energy-BDF1-gspm-2}, the energy evolution over time up to 1 ns is investigated for the unprojected GSPM and BDF1 methods under four different values of $\alpha$ (0.1, 0.05, 0.02, and 0.01). In comparison with the projected schemes, both methods display an increasing divergence in energy behavior as $\alpha$ decreases. For relatively large values of $\alpha$ (0.1 and 0.05), the unprojected GSPM method exhibits considerably stronger energy dissipation than BDF1, and the discrepancy between them enlarges over time. When $\alpha$ becomes smaller (0.02 and particularly 0.01), the unprojected GSPM approach not only presents excessive energy loss but also generates noticeable numerical oscillations at the initial stage, while BDF1 maintains a relatively smooth energy decay. These findings indicate that without the projection operation, the GSPM method cannot maintain the desired energy evolution and suffers from artificial energy dissipation and spurious oscillations, especially for small $\alpha$, whereas BDF1 demonstrates superior stability and robustness over the entire range of $\alpha$ values.

For the dynamics, we proceed the test of domain wall motion.

\begin{figure}[htbp]
    \centering
    \subfloat[Initial state]{\includegraphics[width=0.5\linewidth]{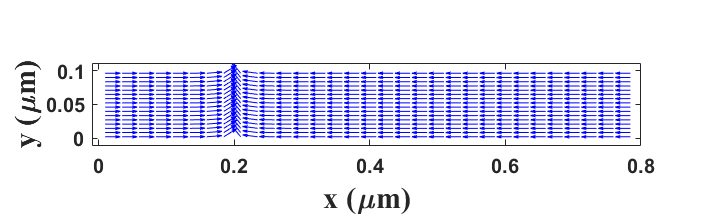}}
    \subfloat[$t=0.1\;ns$]{\includegraphics[width=0.5\linewidth]{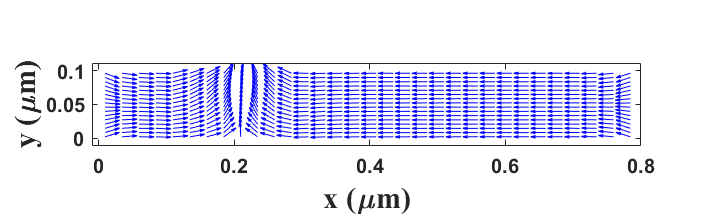}}
    \hspace{0.1in}
     \subfloat[$t=0.5\;ns$]{\includegraphics[width=0.5\linewidth]{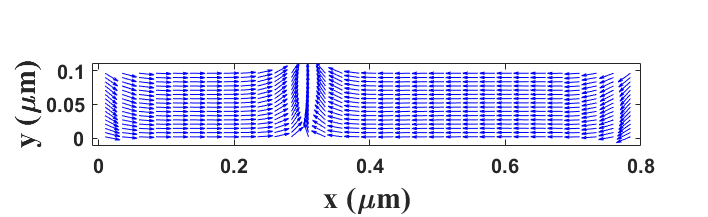}}
    \subfloat[$t=1\;ns$]{\includegraphics[width=0.5\linewidth]{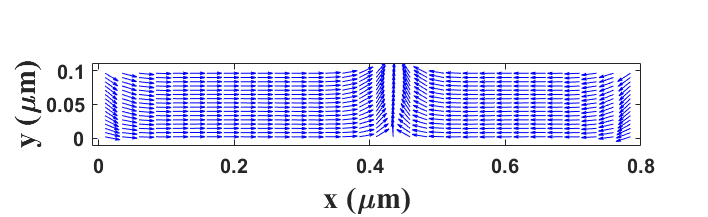}}
  \caption{Domain wall motion using GSPM with projection, $\alpha=0.1$, $He=5\;mT$.}
    \label{fig:motion-1}
\end{figure}

\begin{figure}[htbp]
    \centering
    \subfloat[Initial state]{\includegraphics[width=0.5\linewidth]{motion_he_5mT_alpha_0dot1_intial_projection}}
    \subfloat[$t=0.1\;ns$]{\includegraphics[width=0.5\linewidth]{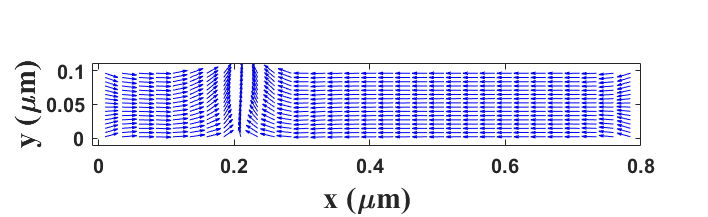}}
    \hspace{0.1in}
     \subfloat[$t=0.5\;ns$]{\includegraphics[width=0.5\linewidth]{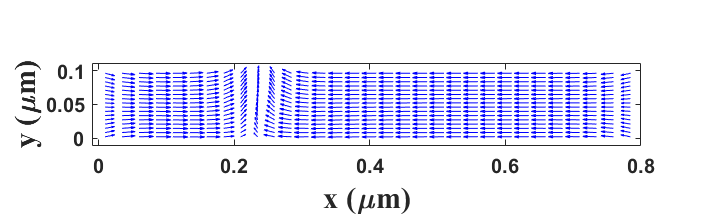}}
    \subfloat[$t=1\;ns$]{\includegraphics[width=0.5\linewidth]{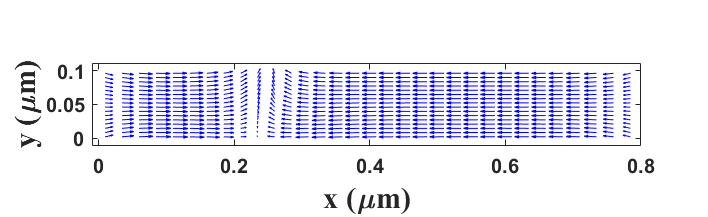}}
  \caption{Domain wall motion using GSPM without projection, $\alpha=0.1$, $He=5\;mT$.}
    \label{fig:motion-2}
\end{figure}

\begin{figure}[htbp]
    \centering
    \subfloat[Initial state]{\includegraphics[width=0.5\linewidth]{motion_he_5mT_alpha_0dot1_intial_projection}}
    \subfloat[$t=0.1\;ns$]{\includegraphics[width=0.5\linewidth]{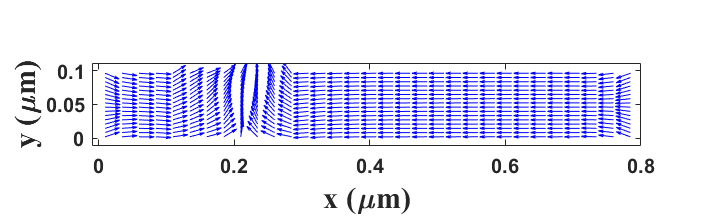}}
    \hspace{0.1in}
     \subfloat[$t=0.5\;ns$]{\includegraphics[width=0.5\linewidth]{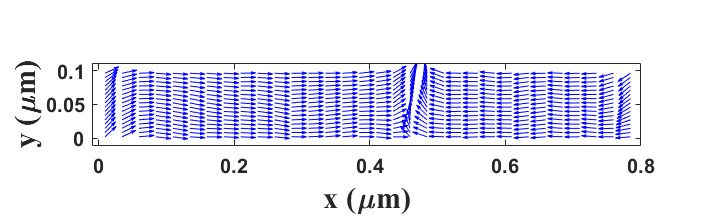}}
    \subfloat[$t=1\;ns$]{\includegraphics[width=0.5\linewidth]{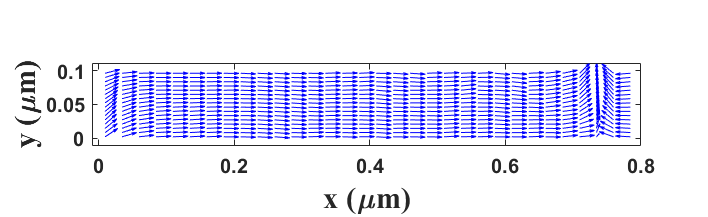}}
  \caption{Domain wall motion using GSPM with projection, $\alpha=0.01$, $He=5\;mT$.}
    \label{fig:motion-3}
\end{figure}

\begin{figure}[htbp]
    \centering
    \subfloat[Initial state]{\includegraphics[width=0.5\linewidth]{motion_he_5mT_alpha_0dot1_intial_projection}}
    \subfloat[$t=0.1\;ns$]{\includegraphics[width=0.5\linewidth]{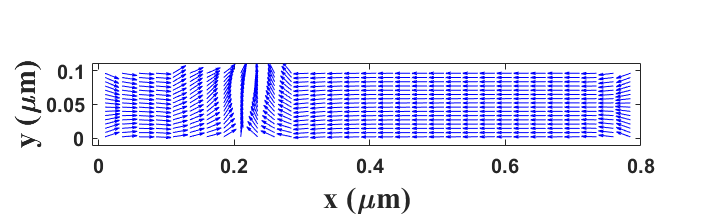}}
    \hspace{0.1in}
     \subfloat[$t=0.5\;ns$]{\includegraphics[width=0.5\linewidth]{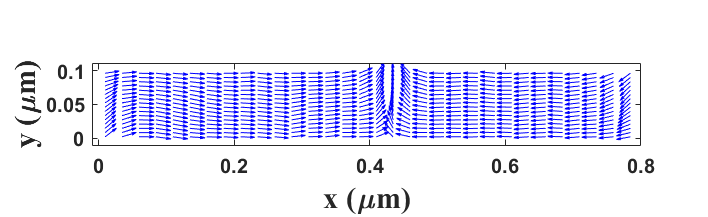}}
    \subfloat[$t=1\;ns$]{\includegraphics[width=0.5\linewidth]{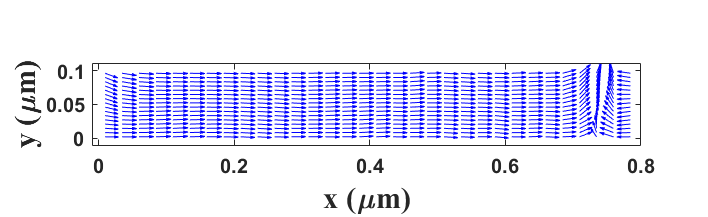}}
  \caption{Domain wall motion using GSPM without projection, $\alpha=0.01$, $He=5\;mT$.}
    \label{fig:motion-4}
\end{figure}

A Ne\'el domain wall was initialized as the initial magnetic state within a ferromagnetic nanostrip of size $800 \times 100 \times 4\,\text{nm}^3$. 
To ensure sufficient spatial resolution for capturing domain wall features while maintaining computational feasibility, the nanostrip was discretized using a structured grid of $128 \times 64 \times 4$ nodes. 
Following initialization, an external magnetic field of magnitude $\boldsymbol{h}_e = 5\,\text{mT}$ was applied along the positive $x$-direction to drive domain wall motion. 
Micromagnetic simulations of domain wall dynamics were performed over a time interval of up to $1\,\text{ns}$. The result for GSPM with projection method given $\alpha=0.1$ is presented in \Cref{fig:motion-1}. The domain wall moves to right driven by the external field. However, the domain wall using GSPM no projection method can not move to the right, which is shown in \Cref{fig:motion-2}. To see the effectiveness, we take $\alpha=0.01$, and get the comparable results using GSPM projection method and GSPM no projection method, which is shown in \Cref{fig:motion-3} and \Cref{fig:motion-4}.

\begin{figure}[htbp]
    \centering
    \subfloat[Initial state]{\includegraphics[width=0.5\linewidth]{motion_he_5mT_alpha_0dot1_intial_projection}}
    \subfloat[$t=0.1\;ns$]{\includegraphics[width=0.5\linewidth]{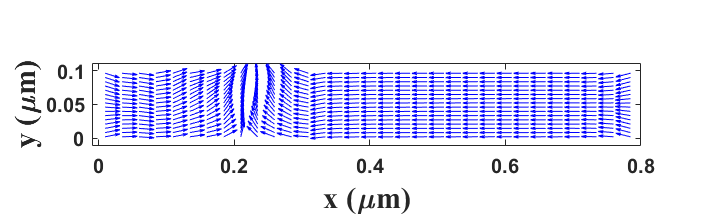}}
    \hspace{0.1in}
     \subfloat[$t=0.5\;ns$]{\includegraphics[width=0.5\linewidth]{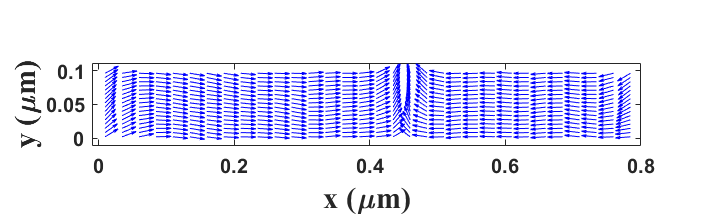}}
    \subfloat[$t=0.8\;ns$]{\includegraphics[width=0.5\linewidth]{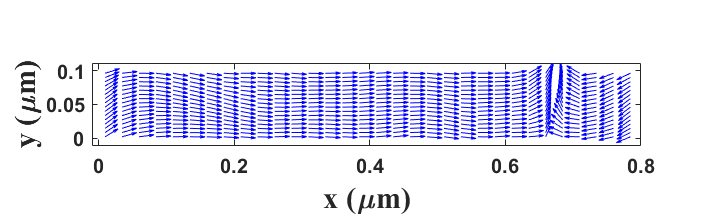}}
  \caption{Domain wall motion using BDF1 with projection, $\alpha=0.01$, $He=5\;mT$.}
    \label{fig:motion-BDF1-1}
\end{figure}

\begin{figure}[htbp]
    \centering
    \subfloat[Initial state]{\includegraphics[width=0.5\linewidth]{motion_he_5mT_alpha_0dot1_intial_projection}}
    \subfloat[$t=0.1\;ns$]{\includegraphics[width=0.5\linewidth]{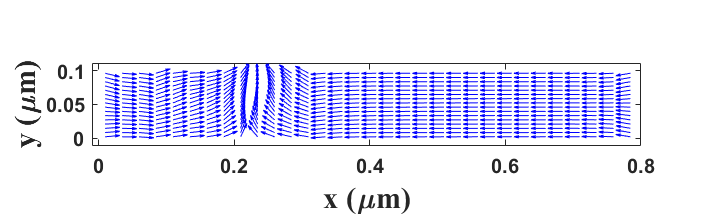}}
    \hspace{0.1in}
     \subfloat[$t=0.5\;ns$]{\includegraphics[width=0.5\linewidth]{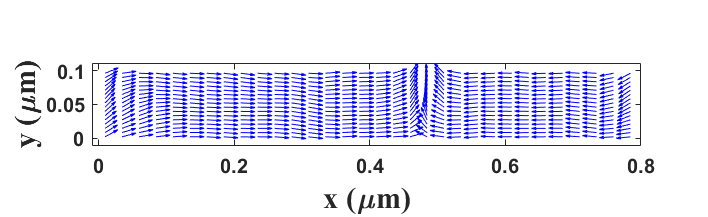}}
    \subfloat[$t=0.8\;ns$]{\includegraphics[width=0.5\linewidth]{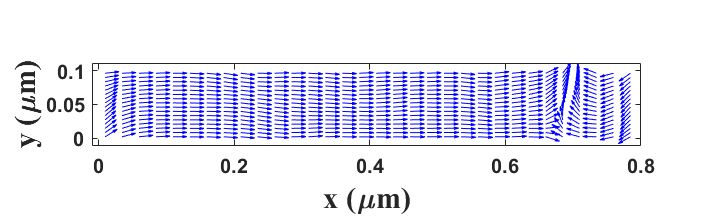}}
  \caption{Domain wall motion using BDF1 without projection, $\alpha=0.01$, $He=5\;mT$.}
    \label{fig:motion-BDF1-2}
\end{figure}

When we apply the BDF1 projection or no projection method to simulate the domain wall motion, we get the results presented in \Cref{fig:motion-BDF1-1} and \Cref{fig:motion-BDF1-2}. In different settings of parameters $\alpha=0.1$ and $\alpha=0.01$, the numerical results of domain wall motion suggest that the BDF1 with projection method is comparable to that of BDF1 no projection method. In this sense, we can use BDF1 without projection method to simulate some micromagnetic settings. 

\begin{figure}[htbp]
    \centering
    \subfloat[Initial state]{\includegraphics[width=0.5\linewidth]{motion_he_5mT_alpha_0dot1_intial_projection}}
    \subfloat[$t=0.1\;ns$]{\includegraphics[width=0.5\linewidth]{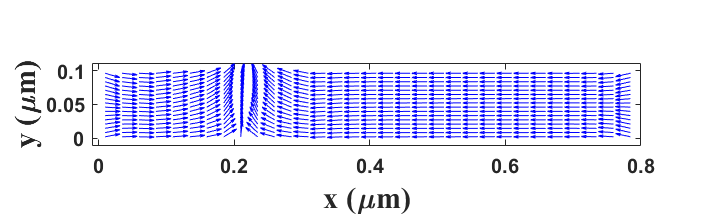}}
    \hspace{0.1in}
     \subfloat[$t=0.5\;ns$]{\includegraphics[width=0.5\linewidth]{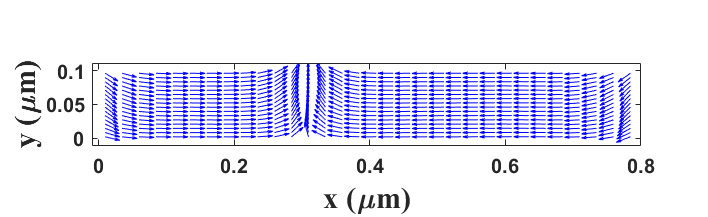}}
    \subfloat[$t=0.8\;ns$]{\includegraphics[width=0.5\linewidth]{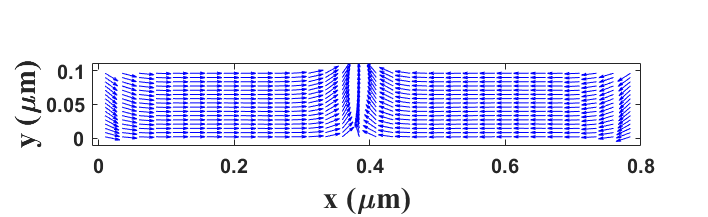}}
  \caption{Domain wall motion using BDF1 with projection, $\alpha=0.1$, $He=5\;mT$.}
    \label{fig:motion-BDF1-4}
\end{figure}

\begin{figure}[htbp]
    \centering
    \subfloat[Initial state]{\includegraphics[width=0.5\linewidth]{motion_he_5mT_alpha_0dot1_intial_projection}}
    \subfloat[$t=0.1\;ns$]{\includegraphics[width=0.5\linewidth]{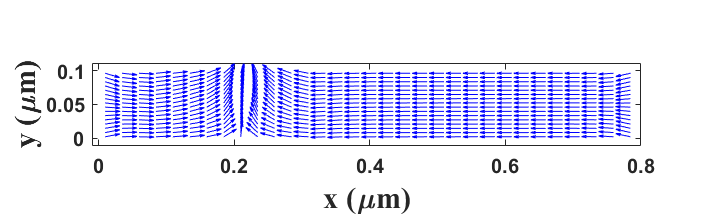}}
    \hspace{0.1in}
     \subfloat[$t=0.5\;ns$]{\includegraphics[width=0.5\linewidth]{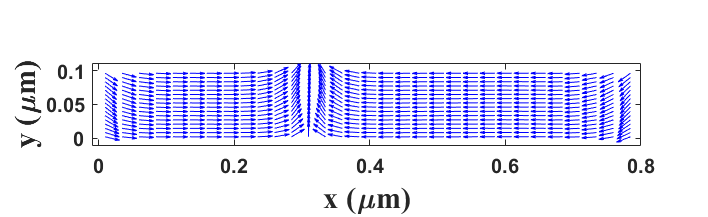}}
    \subfloat[$t=0.8\;ns$]{\includegraphics[width=0.5\linewidth]{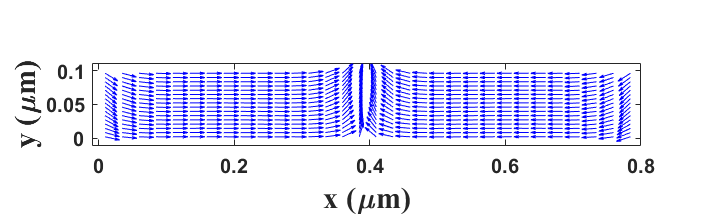}}
  \caption{Domain wall motion using BDF1 without projection, $\alpha=0.1$, $He=5\;mT$.}
    \label{fig:motion-BDF1-3}
\end{figure}


\section{Conclusions and discussions}
\label{sec:conclusions}

In this paper, we compare two types of methods and investigate the effect of the projection step on the actual results of micromagnetic simulations. The results show that in practical simulations, removing the projection step does not bring significant impact on the simulation effect. Our study also indicates that there are obvious differences in the actual performance of the two types of methods: the GSPM method depends on the dissipation coefficient in the calculation results of both the projected and unprojected steps, and under small dissipation coefficients, the results of the GSPM projected method are consistent with those of the unprojected step. The BDF1 method reported in this paper does not depend on the dissipation coefficient in the calculation results of the projected and unprojected steps, and the results of the BDF1 projected method are consistent with those of the unprojected step.

In this paper, we only carry out applied research on the first-order projected and unprojected methods, and the research results on the projected and unprojected high-order methods can be similarly extended. In addition, mathematically, projecting onto a sphere is a nonlinear operation, which brings essential difficulties to the numerical analysis of the projection method; mathematically, the unprojected method is easier to analyze. This paper suggests that the application effect of the unprojected method will not be much worse than that of the projection method.

\section*{Acknowledgments}
This work is supported in part by the Jiangsu Science and Technology Programme-Fundamental Research Plan Fund (BK20250468), and the Research and Development Fund of Xi'an Jiaotong Liverpool University (RDF-24-01-015).

\appendix

\section{GSPM with or without projection}
\begin{figure}[htbp]
    \centering
    \subfloat[Initial C state arrow]{\includegraphics[width=0.3\linewidth]{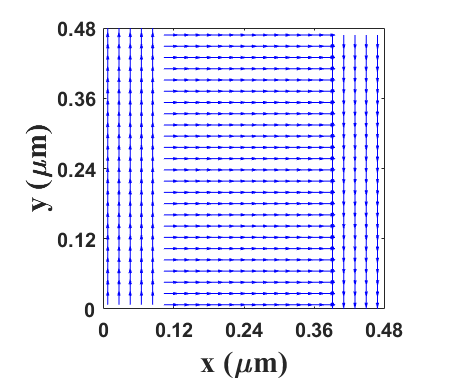}}
     \subfloat[GSPM projection arrow]{\includegraphics[width=0.3\linewidth]{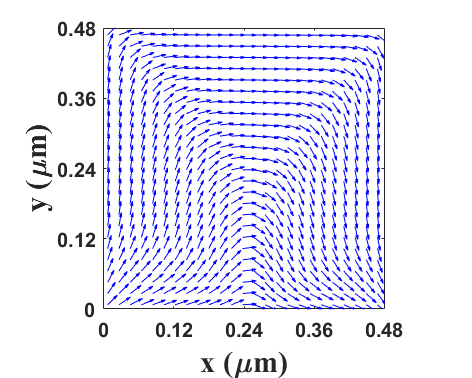}}
     \subfloat[GSPM no projection arrow]{\includegraphics[width=0.3\linewidth]{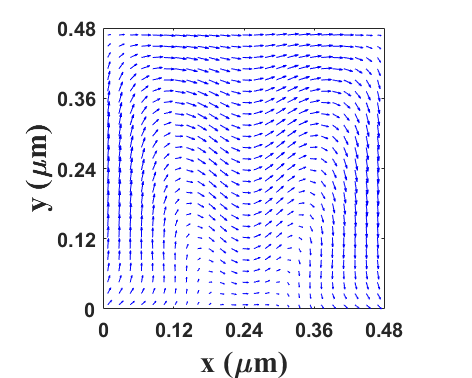}}
    \hspace{0.1in}
     \subfloat[Initial C state angle]{\includegraphics[width=0.3\linewidth]{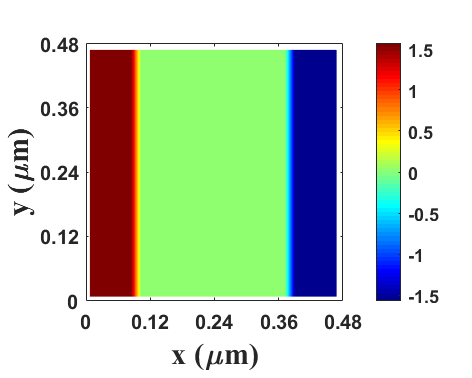}}
      \subfloat[GSPM projection angle]{\includegraphics[width=0.3\linewidth]{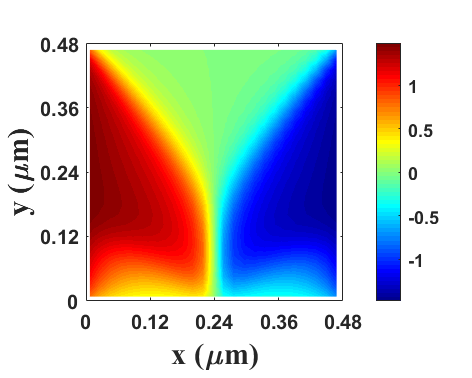}}
      \subfloat[GSPM noprojection angle]{\includegraphics[width=0.3\linewidth]{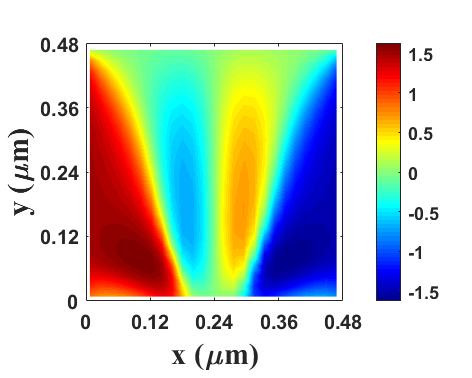}}
      \hspace{0.1in}
       \subfloat[Initial Diamond state arrow]{\includegraphics[width=0.3\linewidth]{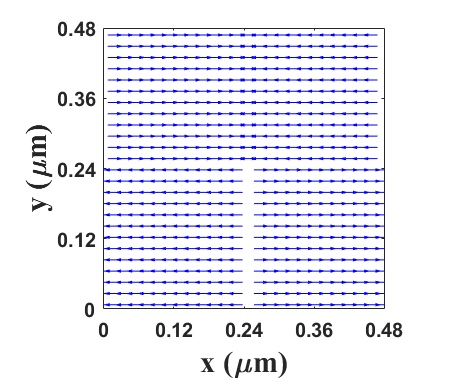}}
     \subfloat[GSPM projection arrow]{\includegraphics[width=0.3\linewidth]{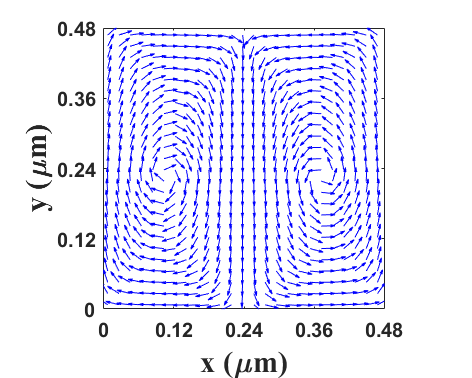}}
     \subfloat[GSPM no projection arrow]{\includegraphics[width=0.3\linewidth]{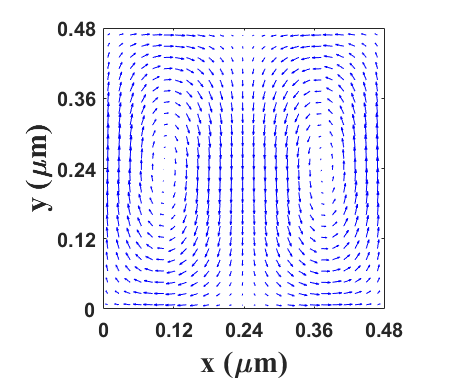}}
     \hspace{0.1in}
     \subfloat[Initial Diamond state angle]{\includegraphics[width=0.3\linewidth]{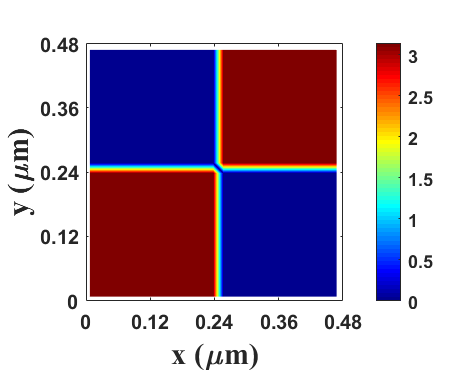}}
      \subfloat[GSPM projection angle]{\includegraphics[width=0.3\linewidth]{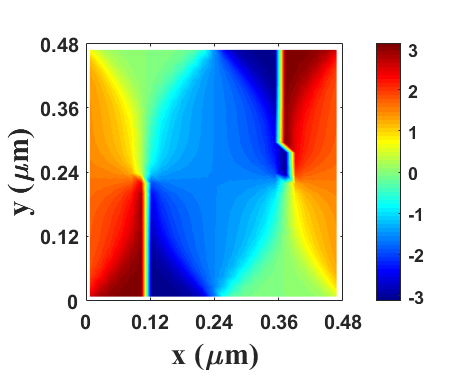}}
      \subfloat[GSPM noprojection angle]{\includegraphics[width=0.3\linewidth]{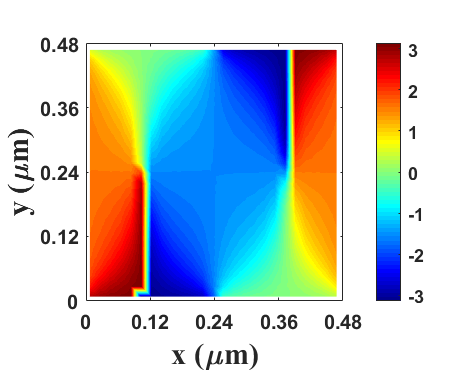}}
   \caption{The setup $\alpha=0.1$ and the final time $1\;ns$. Initial state is given by left panel. The results by GSPM projection in middle panel. The results by GSPM no projection in middle panel.}
    \label{fig:2}
\end{figure}

\begin{figure}[htbp]
    \centering
    \subfloat[Initial Flower state arrow]{\includegraphics[width=0.3\linewidth]{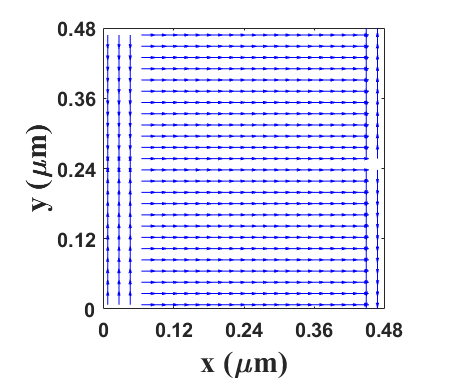}}
     \subfloat[GSPM projection arrow]{\includegraphics[width=0.3\linewidth]{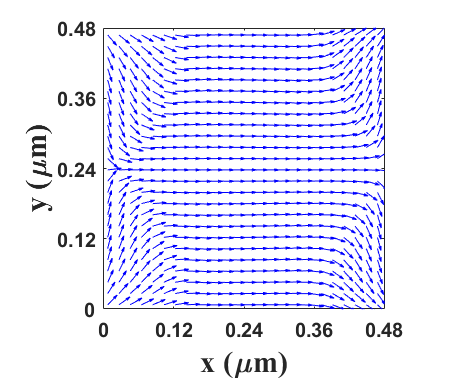}}
     \subfloat[GSPM no projection arrow]{\includegraphics[width=0.3\linewidth]{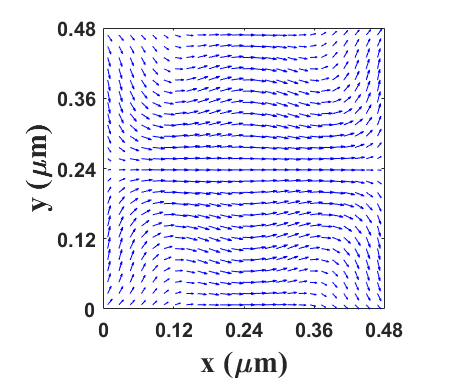}}
    \hspace{0.1in}
     \subfloat[Initial Flower state angle]{\includegraphics[width=0.3\linewidth]{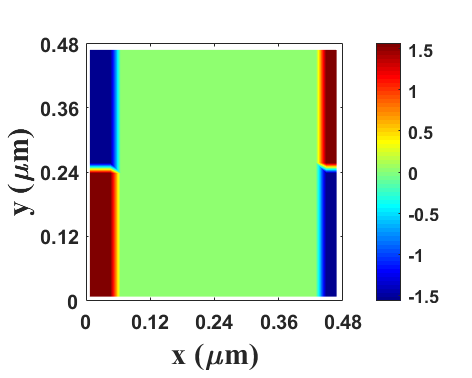}}
      \subfloat[GSPM projection angle]{\includegraphics[width=0.3\linewidth]{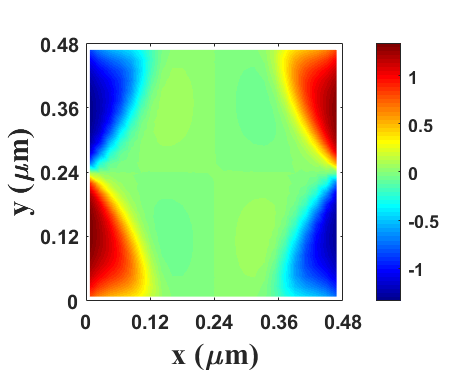}}
      \subfloat[GSPM noprojection angle]{\includegraphics[width=0.3\linewidth]{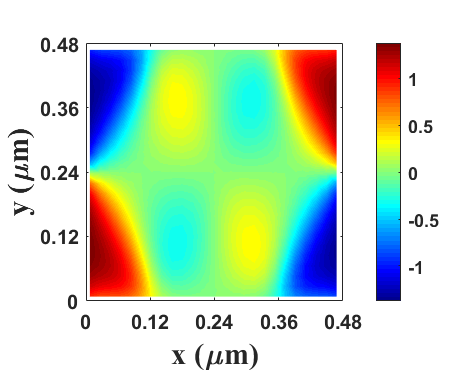}}
      \hspace{0.1in}
       \subfloat[Initial random state arrow]{\includegraphics[width=0.3\linewidth]{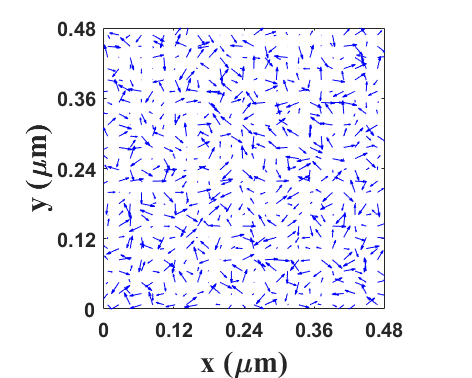}}
     \subfloat[GSPM projection arrow]{\includegraphics[width=0.3\linewidth]{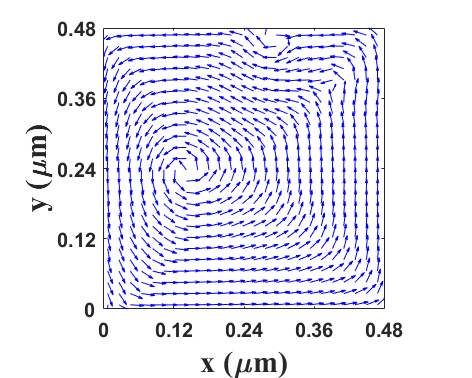}}
     \subfloat[GSPM no projection arrow]{\includegraphics[width=0.3\linewidth]{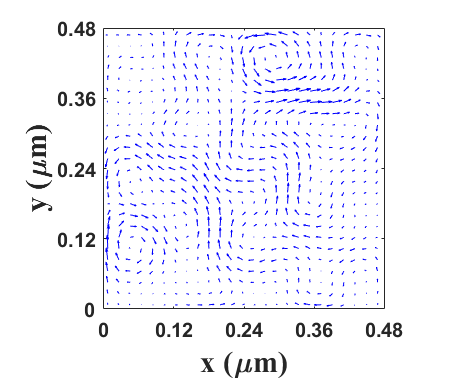}}
     \hspace{0.1in}
     \subfloat[Initial random state angle]{\includegraphics[width=0.3\linewidth]{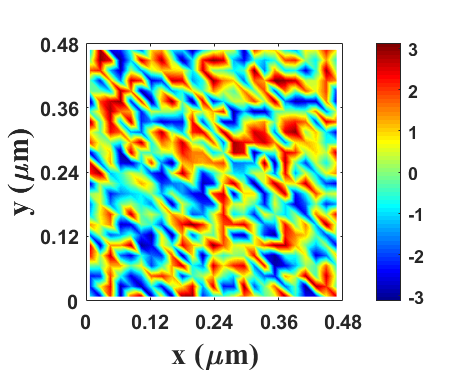}}
      \subfloat[GSPM projection angle]{\includegraphics[width=0.3\linewidth]{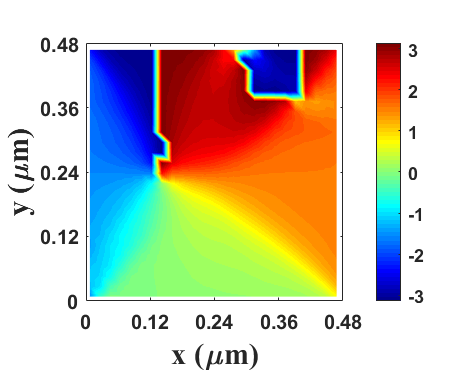}}
      \subfloat[GSPM noprojection angle]{\includegraphics[width=0.3\linewidth]{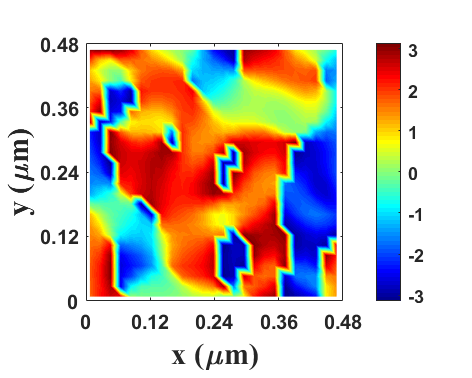}}
    \caption{The setup $\alpha=0.1$ and the final time $1\;ns$. Initial state is given by left panel. The results by GSPM projection in middle panel. The results by GSPM no projection in middle panel.}
    \label{fig:3}
\end{figure}

\begin{figure}[htbp]
    \centering
    \subfloat[Initial Single Crosstie state arrow]{\includegraphics[width=0.3\linewidth]{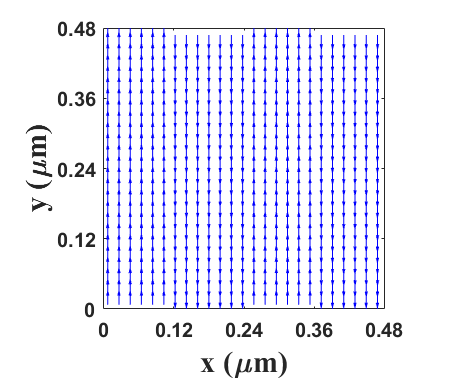}}
     \subfloat[GSPM projection arrow]{\includegraphics[width=0.3\linewidth]{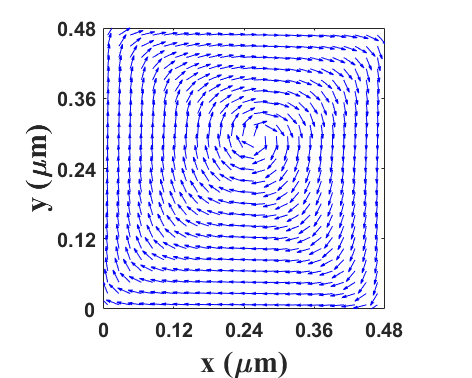}}
     \subfloat[GSPM no projection arrow]{\includegraphics[width=0.3\linewidth]{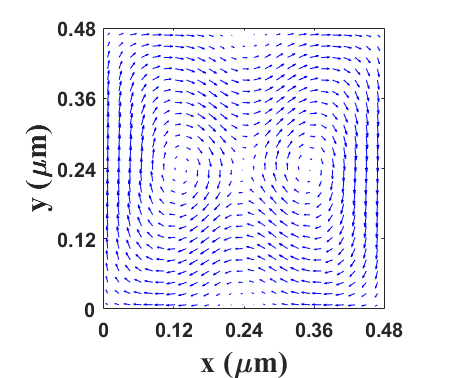}}
    \hspace{0.1in}
     \subfloat[Initial Single Crosstie state angle]{\includegraphics[width=0.3\linewidth]{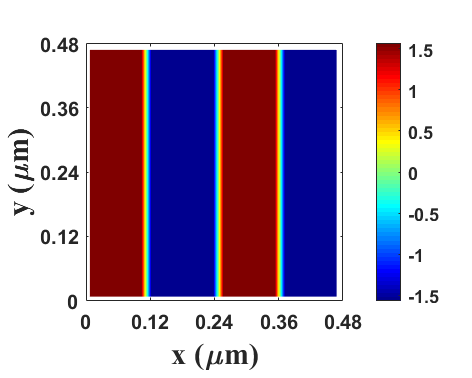}}
      \subfloat[GSPM projection angle]{\includegraphics[width=0.3\linewidth]{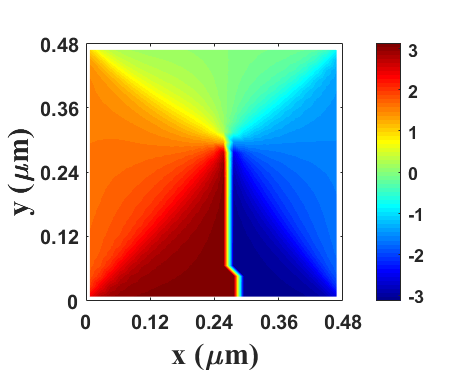}}
      \subfloat[GSPM noprojection angle]{\includegraphics[width=0.3\linewidth]{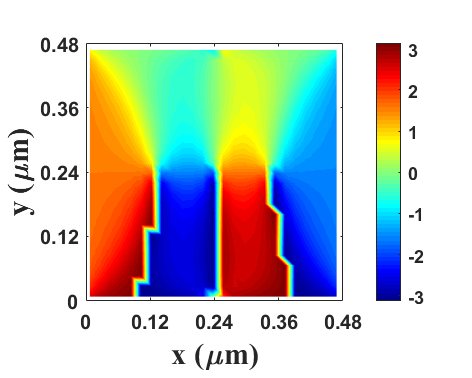}}
      \hspace{0.1in}
       \subfloat[Initial Double Crosstie state arrow]{\includegraphics[width=0.3\linewidth]{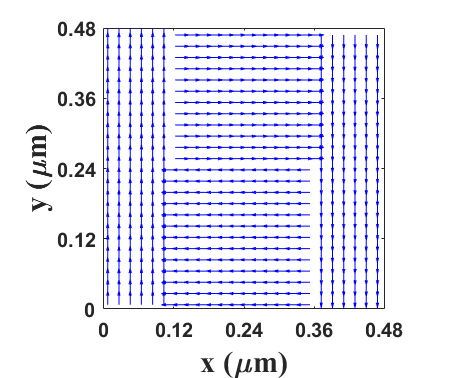}}
     \subfloat[GSPM projection arrow]{\includegraphics[width=0.3\linewidth]{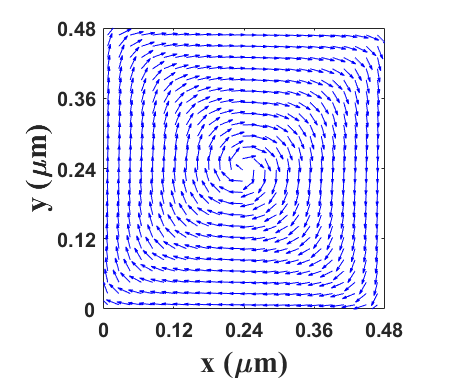}}
     \subfloat[GSPM no projection arrow]{\includegraphics[width=0.3\linewidth]{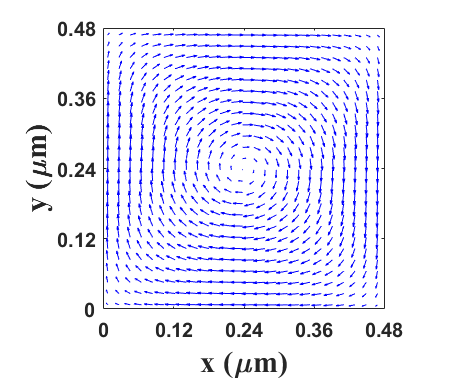}}
     \hspace{0.1in}
     \subfloat[Initial Double Crosstie state angle]{\includegraphics[width=0.3\linewidth]{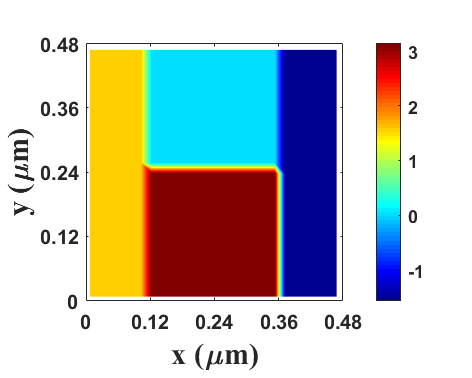}}
      \subfloat[GSPM projection angle]{\includegraphics[width=0.3\linewidth]{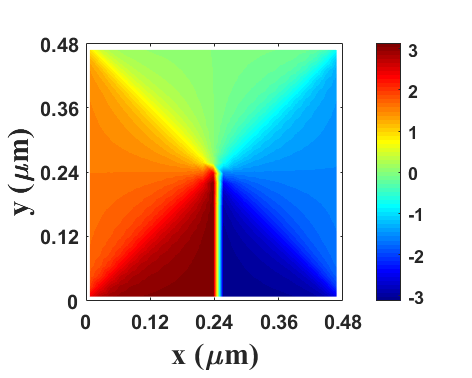}}
      \subfloat[GSPM noprojection angle]{\includegraphics[width=0.3\linewidth]{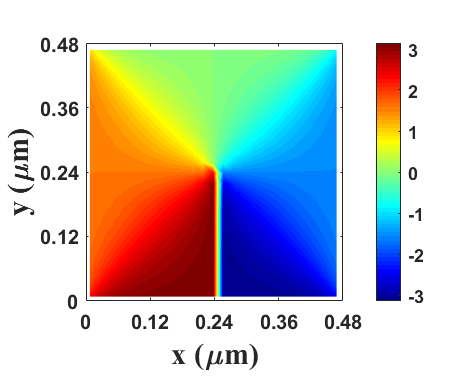}}
   \caption{The setup $\alpha=0.1$ and the final time $1\;ns$. Initial state is given by left panel. The results by GSPM projection in middle panel. The results by GSPM no projection in middle panel.}
    \label{fig:4}
\end{figure}

\section{BDF1 with or without projection}
\begin{figure}[htbp]
    \centering
    \subfloat[Initial C state arrow]{\includegraphics[width=0.3\linewidth]{gspm_noprojection_arrow_C_initial}}
     \subfloat[BDF1 projection arrow]{\includegraphics[width=0.3\linewidth]{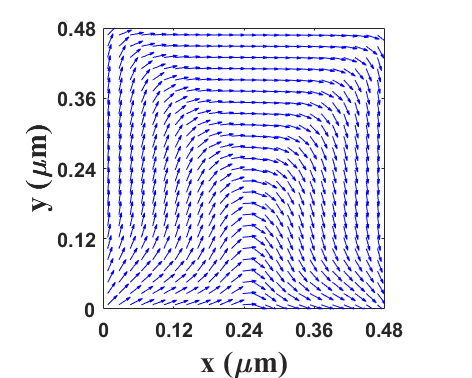}}
     \subfloat[BDF1 no projection arrow]{\includegraphics[width=0.3\linewidth]{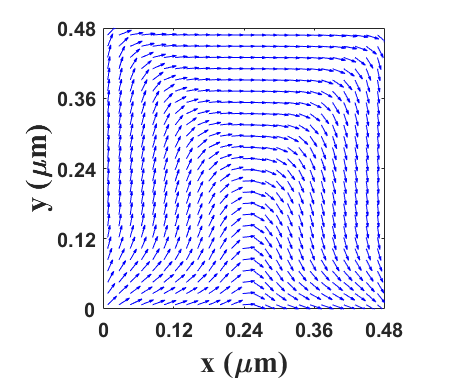}}
    \hspace{0.1in}
     \subfloat[Initial C state angle]{\includegraphics[width=0.3\linewidth]{gspm_noprojection_color_C_initial}}
      \subfloat[BDF1 projection angle]{\includegraphics[width=0.3\linewidth]{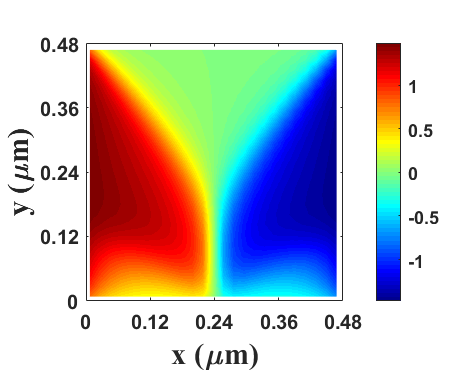}}
      \subfloat[BDF1 noprojection angle]{\includegraphics[width=0.3\linewidth]{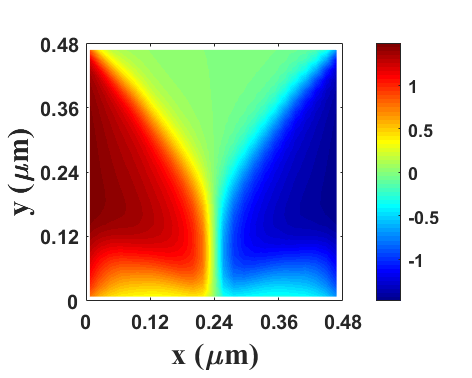}}
      \hspace{0.1in}
       \subfloat[Initial Diamond state arrow]{\includegraphics[width=0.3\linewidth]{gspm_noprojection_arrow_Diamond_initial}}
     \subfloat[BDF1 projection arrow]{\includegraphics[width=0.3\linewidth]{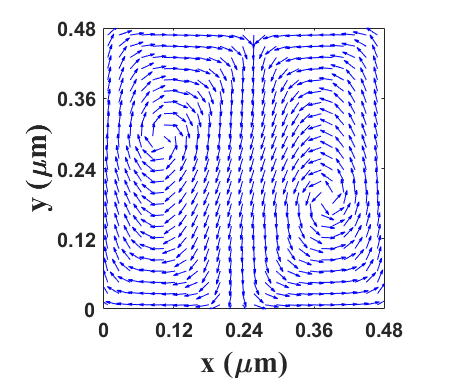}}
     \subfloat[BDF1 no projection arrow]{\includegraphics[width=0.3\linewidth]{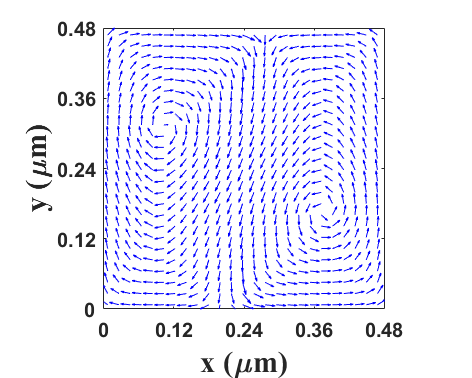}}
     \hspace{0.1in}
     \subfloat[Initial Diamond state angle]{\includegraphics[width=0.3\linewidth]{gspm_noprojection_color_Diamond_initial}}
      \subfloat[BDF1 projection angle]{\includegraphics[width=0.3\linewidth]{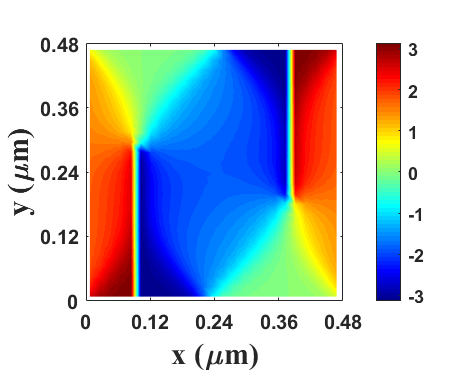}}
      \subfloat[BDF1 noprojection angle]{\includegraphics[width=0.3\linewidth]{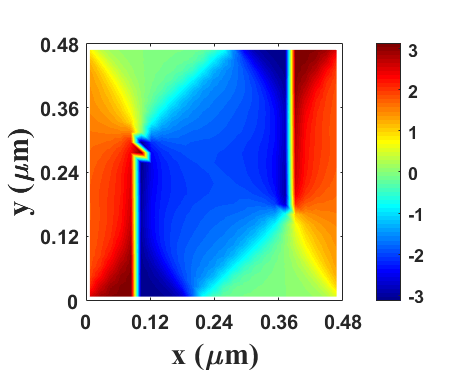}}
   \caption{The setup $\alpha=0.1$ and the final time $1\;ns$. Initial state is given by left panel. The results by BDF1 projection in middle panel. The results by BDF1 no projection in right panel.}
    \label{fig:BDF1-2}
\end{figure}

\begin{figure}[htbp]
    \centering
    \subfloat[Initial Flower state arrow]{\includegraphics[width=0.3\linewidth]{gspm_noprojection_arrow_Flower_initial}}
     \subfloat[BDF1 projection arrow]{\includegraphics[width=0.3\linewidth]{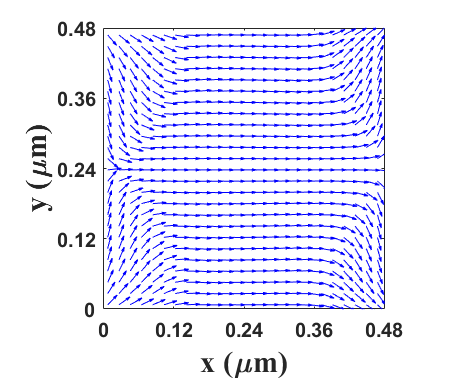}}
     \subfloat[BDF1 no projection arrow]{\includegraphics[width=0.3\linewidth]{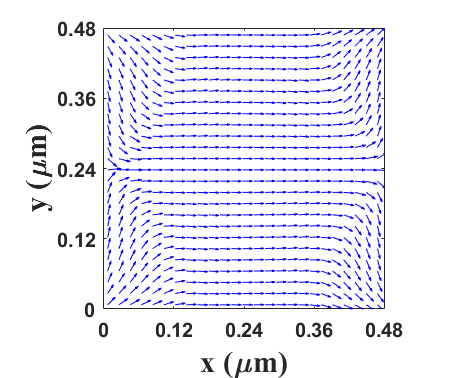}}
    \hspace{0.1in}
     \subfloat[Initial Flower state angle]{\includegraphics[width=0.3\linewidth]{gspm_noprojection_color_Flower_initial}}
      \subfloat[BDF1 projection angle]{\includegraphics[width=0.3\linewidth]{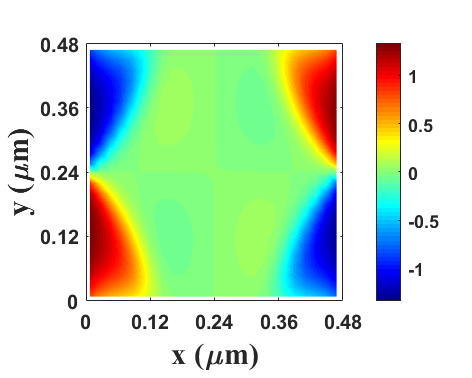}}
      \subfloat[BDF1 noprojection angle]{\includegraphics[width=0.3\linewidth]{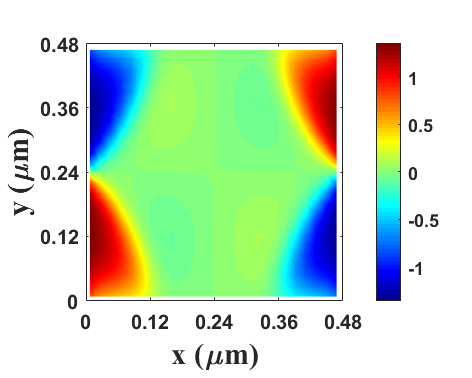}}
      \hspace{0.1in}
       \subfloat[Initial random state arrow]{\includegraphics[width=0.3\linewidth]{gspm_noprojection_arrow_random_initial}}
     \subfloat[BDF1 projection arrow]{\includegraphics[width=0.3\linewidth]{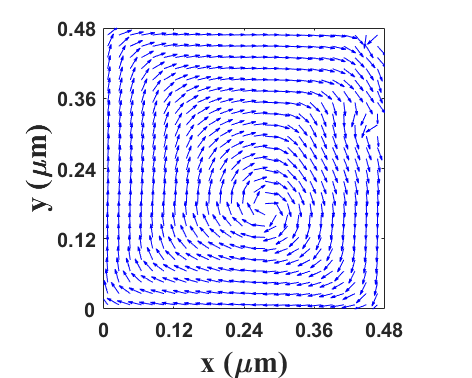}}
     \subfloat[BDF1 no projection arrow]{\includegraphics[width=0.3\linewidth]{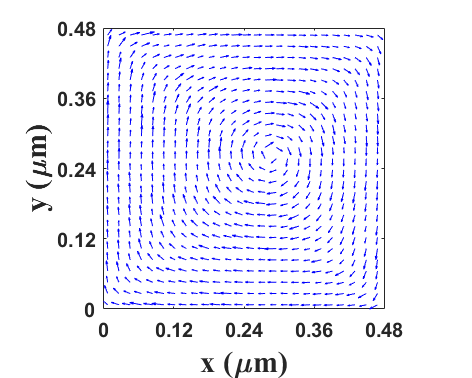}}
     \hspace{0.1in}
     \subfloat[Initial random state angle]{\includegraphics[width=0.3\linewidth]{gspm_noprojection_color_random_initial}}
      \subfloat[BDF1 projection angle]{\includegraphics[width=0.3\linewidth]{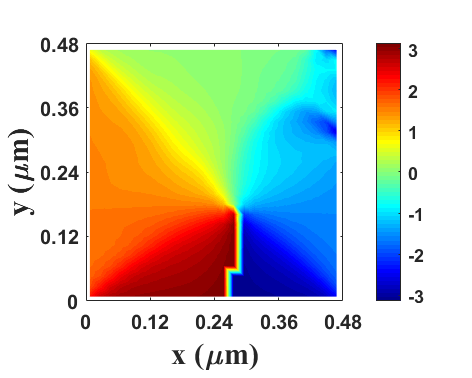}}
      \subfloat[BDF1 noprojection angle]{\includegraphics[width=0.3\linewidth]{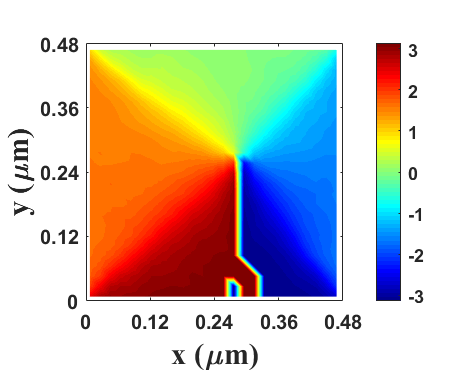}}
    \caption{The setup $\alpha=0.1$ and the final time $1\;ns$. Initial state is given by left panel. The results by BDF1 projection in middle panel. The results by BDF1 no projection in right panel.}
    \label{fig:BDF1-3}
\end{figure}

\begin{figure}[htbp]
    \centering
    \subfloat[Initial Single Crosstie state arrow]{\includegraphics[width=0.3\linewidth]{gspm_noprojection_arrow_SingleCrosstie_initial}}
     \subfloat[BDF1 projection arrow]{\includegraphics[width=0.3\linewidth]{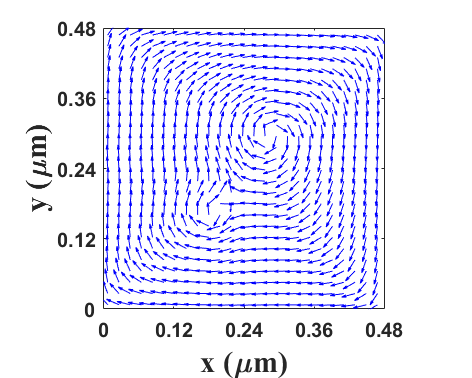}}
     \subfloat[BDF1 no projection arrow]{\includegraphics[width=0.3\linewidth]{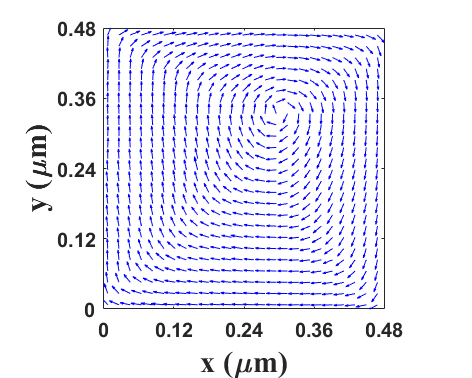}}
    \hspace{0.1in}
     \subfloat[Initial Single Crosstie state angle]{\includegraphics[width=0.3\linewidth]{gspm_noprojection_color_SingleCrosstie_initial}}
      \subfloat[BDF1 projection angle]{\includegraphics[width=0.3\linewidth]{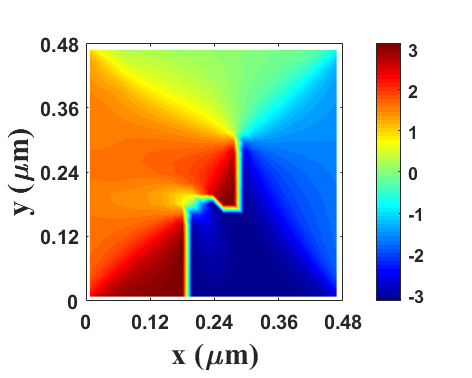}}
      \subfloat[BDF1 noprojection angle]{\includegraphics[width=0.3\linewidth]{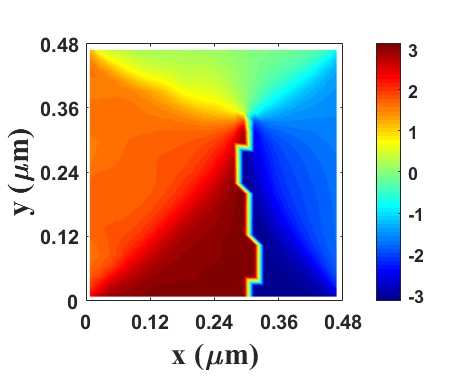}}
      \hspace{0.1in}
       \subfloat[Initial Double Crosstie state arrow]{\includegraphics[width=0.3\linewidth]{gspm_noprojection_arrow_DoubleCrosstie_initial}}
     \subfloat[BDF1 projection arrow]{\includegraphics[width=0.3\linewidth]{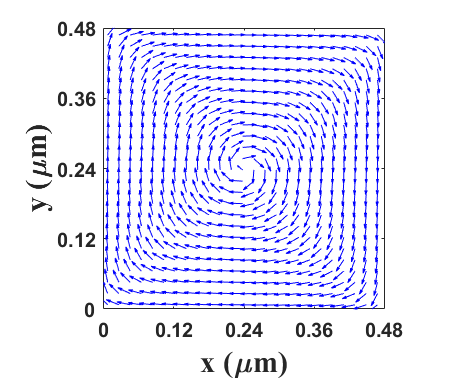}}
     \subfloat[BDF1 no projection arrow]{\includegraphics[width=0.3\linewidth]{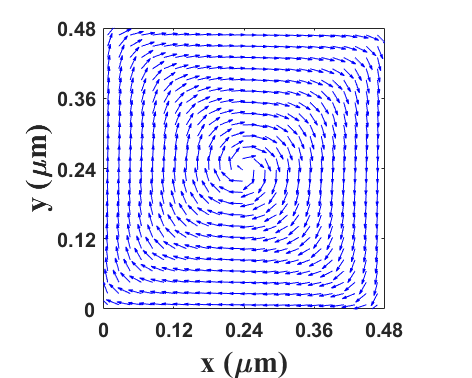}}
     \hspace{0.1in}
     \subfloat[Initial Double Crosstie state angle]{\includegraphics[width=0.3\linewidth]{gspm_noprojection_color_DoubleCrosstie_initial}}
      \subfloat[BDF1 projection angle]{\includegraphics[width=0.3\linewidth]{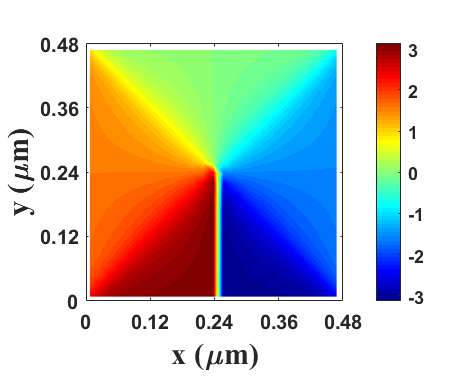}}
      \subfloat[BDF1 no projection angle]{\includegraphics[width=0.3\linewidth]{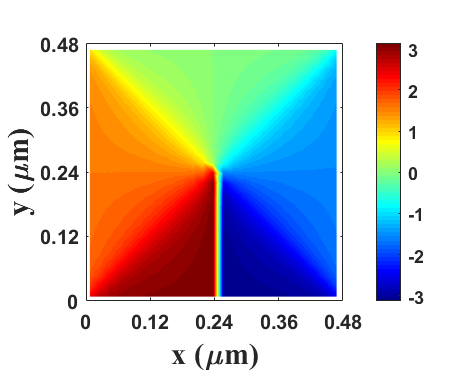}}
   \caption{The setup $\alpha=0.1$ and the final time $1\;ns$. Initial state is given by left panel. The results by BDF1 projection in middle panel. The results by BDF1 no projection in right panel.}
    \label{fig:BDF1-4}
\end{figure}

\vspace{1cm}

\bibliographystyle{elsarticle-num-names}
\bibliography{references.bib}

\end{document}